\documentclass[11pt]{article}
\usepackage{graphicx}
\usepackage{latexsym}
\usepackage[left=1in,top=1in,right=1in,bottom=1in]{geometry}
\usepackage{amsmath}
\usepackage{booktabs}
\usepackage{color}
\usepackage{tikz}
\usepackage{graphicx}
\graphicspath{{figures/}}
\usepackage{subcaption}
\usepackage{amssymb}
\usepackage{graphics}
\usepackage{amsthm}
\usepackage{float}
\usepackage{cite}
\usepackage{longtable}
\usepackage{amsmath}
\usepackage{algorithm} 
\usepackage{algpseudocode}
\usepackage{algorithmicx}
\usepackage{natbib}

\DeclareMathOperator{\E}{\mathbb{E}}

\DeclareMathOperator*{\argmin}{arg\,min}
\DeclareMathOperator*{\logit}{logit}

\newcommand{\trc}[1]{\textcolor{black}{#1}}
\newcommand{\at}[2][]{#1|_{#2}}

\usepackage{hyperref}
\usetikzlibrary{shapes}
\usetikzlibrary{arrows}
\definecolor{darkblue}{rgb}{0,0.4,0.9}
\definecolor{gray10}{rgb}{0.1,0.1,0.1}
\definecolor{gray20}{rgb}{0.2,0.2,0.2}
\definecolor{gray30}{rgb}{0.3,0.3,0.3}
\definecolor{gray40}{rgb}{0.4,0.4,0.4}
\definecolor{gray60}{rgb}{0.6,0.6,0.6}
\definecolor{gray80}{rgb}{0.8,0.8,0.8}
\definecolor{gray90}{rgb}{0.9,0.9,.9}
\definecolor{gray95}{rgb}{0.95,0.95,.95}
\definecolor{gray96}{rgb}{0.96,0.96,.96}
\definecolor{lgreen} {RGB}{180,210,100}
\definecolor{dblue}  {RGB}{20,66,129}
\definecolor{ddblue} {RGB}{11,36,69}
\definecolor{lred}   {RGB}{220,0,0}
\definecolor{nred}   {RGB}{224,0,0}
\definecolor{norange}{RGB}{230,120,20}
\definecolor{nyellow}{RGB}{255,221,0}
\definecolor{ngreen} {RGB}{98,158,31}
\definecolor{dgreen} {RGB}{78,138,21}
\definecolor{nblue}  {RGB}{28,130,185}
\definecolor{jblue}  {RGB}{20,50,100}
\definecolor{nnyellow}{RGB}{235,200,0}
\definecolor{purple}{RGB}{150, 0, 120}
\definecolor{sgGreen} {RGB}{20, 180, 50}
\definecolor{revised}{rgb}{0,0,0.9}

\newtheorem{assumption}{Assumption}

\newcommand{\openr}{\hbox{${\rm I\kern-.2em R}$}}
\newcommand{\openn}{\hbox{${\rm I\kern-.2em N}$}}
\newcommand{\indep}{\rotatebox[origin=c]{90}{$\models$}}
\makeatletter
\newcommand*{\defeq}{\mathrel{\rlap{%
      \raisebox{0.3ex}{$\m@th\cdot$}}%
    \raisebox{-0.3ex}{$\m@th\cdot$}}%
  =}
\makeatother

\begin{document}

\title{Collaborative-controlled LASSO  for Constructing Propensity Score-based Estimators in High-Dimensional Data}

\author{Cheng Ju $^1$,  Richard Wyss$^2$, Jessica M. Franklin$^2$, \\Sebastian Schneeweiss$^2$, Jenny H{\"a}ggstr{\"o}m $^3$,\\ Mark J. van der Laan $^1$}

\date{%
   $^1$ Division of Biostatistics, University of California, Berkeley\\
   $^2$ Division of Pharmacoepidemiology and Pharmacoeconomics, Department of Medicine, Brigham and Women’s Hospital and Harvard Medical School\\
   $^3$ Department of Statistics, USBE, Ume{\aa}  University
}

\maketitle
\begin{abstract}
Propensity score (PS) based estimators are increasingly used for causal inference in observational studies. However, model selection for PS estimation in high-dimensional data has received little attention. In these settings, PS models have traditionally been selected based on the goodness-of-fit for the treatment mechanism itself, without consideration of the causal parameter of interest. Collaborative minimum loss-based estimation (C-TMLE) is a novel methodology for causal inference that takes into account information on the causal parameter of interest when selecting a PS model. This ``collaborative learning'' considers variable associations with both treatment and outcome when selecting a PS model in order to minimize a bias-variance trade off in the estimated treatment effect. In this study, we introduce a novel approach for collaborative model selection when using the LASSO estimator for PS estimation  in high-dimensional covariate settings. To demonstrate the importance of selecting the PS model collaboratively, we designed quasi-experiments based on a real electronic healthcare  database, where only the potential outcomes were manually generated, and the treatment and baseline covariates remained unchanged.
Results showed that the C-TMLE algorithm outperformed other competing estimators for both point estimation and confidence interval coverage. In addition, the PS model selected by C-TMLE could be applied to other PS-based estimators, which also resulted in substantive improvement for both point estimation and confidence interval coverage. We illustrate the discussed concepts through an empirical example comparing the effects of non-selective nonsteroidal anti-inflammatory drugs with selective COX-2 inhibitors on gastrointestinal complications in a population of Medicare beneficiaries.  
\end{abstract}

\noindent \textbf{Keywords:}  Propensity Score; Average Treatment Effect; LASSO; Model Selection; Electronic Healthcare Database; Collaborative Targeted Minimum Loss-based Estimation;

\section{Introduction}

\subsection{Purpose}

The propensity score (PS) is defined as the conditional probability of treatment assignment, given a set of pre-treatment covariates \citep{rosenbaum1983central,imbens2000role}. The PS, which we will denote as $g_0$, is widely used to control for confounding  bias  in observational studies. In practice, the PS is usually unknown and PS based estimators must rely on an estimate  of the PS, which we will denote as $g_n$.

Accurately modeling and assessing the validity of fitted PS models is crucial for all PS-based methods. It is generally recommended that PS models be validated through measures of covariate balance across treatment groups after PS adjustment. In high-dimensional covariate settings, however, evaluating covariate balance on very large numbers of variables can be difficult. Using covariate balance to validate PS models in high-dimensional covariate settings is further complicated when applying machine learning algorithms and penalized regression methods to reduce the dimension of the covariate set, as it is not always clear on what variables balance should be evaluated. Cross-validated prediction diagnostics can greatly simplify validation of the PS model when applying machine learning algorithms for PS estimation in high-dimensional covariate settings.

\citep{westreich2010propensity} suggested that machine learning (ML) methods (e.g. support vector machines) could enhance the validity of propensity score estimation, and that ``external'' cross-validation (CV) can be used for model selection. \citep{lee2010improving} further investigated PS weighted estimators when the PS was estimated by multiple ML algorithms, where the hyper-parameters of the ML algorithms were selected by minimizing the CV loss for treatment prediction. Estimation procedures that are based on external CV will result in estimated models that optimize the bias-variance tradeoff for treatment prediction (i.e., the true PS function), but they do not consider the ultimate goal of optimizing the bias-variance tradeoff for the treatment effect estimate. We conjecture that PS estimators that are selected by CV will tend to be over-smoothed in order to reduce variability in the prediction of treatment assignment, and that the optimal estimator in reducing bias in the estimated treatment effect should be less smooth compared to the estimator selected by  external CV. 

To address this limitation of  external CV, we studied two recently proposed variations of the C-TMLE algorithm \citep{ju2017ctmle,van2017ctmle}, and compared them to other widely used estimators using multiple simulation studies. We focused on strategies that combined the C-TMLE algorithms with LASSO regression, an $l\text{-}1$ regularized logistic regression \citep{tibshirani1996regression}, for PS estimation.  Previous studies have shown that LASSO regression can perform well for variable selection when estimating high-dimensional PSs \citep{franklin2015regularized}. However, selecting the optimal tuning parameters to optimize confounding control remains challenging. Combining variations of the C-TMLE algorithm with LASSO regression provides a robust data adaptive approach to PS model selection in high-dimensional covariate datasets, but remains untested. We used quasi-experiments based on a real empirical dataset to  evaluate the performance of combining variations of the C-TMLE algorithm with LASSO regression and demonstrate that exernal CV for model selection is insufficient.

The article is organized as follows. In section \ref{sec:begin}, we introduce the structure of the observed data, the scientific question, the parameter of interest, the average treatment effect (ATE), and the necessary assumptions for making the parameter of interest identifiable. In section \ref{sec:review} we briefly review some commonly used estimators of the ATE. In section \ref{sec:tmle}, we review the targeted minimum-loss based estimator. In section \ref{sec:ctmle}, we introduce two recently proposed C-TMLE algorithms which extend the vanilla TMLE algorithm. In section \ref{sec:data} we describe the electronic healthcare database used in the simulations and empirical analyses. In section \ref{sec:quasi} we describe how the simulated data are generated from the empirical dataset, and how results were analyzed from the simulation, including point estimation (subsection \ref{subsec:point_est}), confidence interval (subsection \ref{subsec:ci}), and pair-wise comparisons (subsection \ref{subsec:pair}) of estimators. In section \ref{sec:da} we apply the vanilla TMLE and novel C-TMLE algorithms to analyze the empirical dataset. Finally, in section \ref{sec:diss}, we discuss the results from the simulations and the scientific findings from the empirical data analysis.

\subsection{Data Structure, Scientific Question, and Identification}
\label{sec:begin}

Suppose we observe $n$ independent and identically distributed (i.i.d.) observations, $O_i=(Y_i,A_i, W_i)$, $i \in 1, \ldots, n$, from some unknown but fixed data generating distribution $P_0$. Consider a simple setting, where $W_i$ is a vector of some pre-treatment baseline covariates of the $i$-th observation, and $A_i$ is a binary indicator taking on a value of 1 if observation $i$ is in the treatment group and is 0 otherwise. Further, suppose that each observation has a counterfactual outcome pair, $(Y_{i0},Y_{i1})$, corresponding to the potential outcome if patient $i$ is in the control group ($A_i = 0$) or the treatment group ($A_i = 1)$. Thus, for each observation, we only observe one of the potential outcomes, $Y_i$, which corresponds with either $Y_{i0}$ or $Y_{i1}$, depending on whether the individual received treatment or remained untreated. For simplicity, we refer to $Q_0(W)$ as the marginal distribution of $W$,  $g_0(W)$ as the conditional expectation of $A|W$, and $\bar{Q}_0(A,W)$ as the conditional expectation of $Y|A,W$. We will let $g_0$ represent the PS, under the data generating distribution $P_0$. In addition, we will let $\E_0$ represent the expectation under the unknown true data generating distribution $P_0$. Consider the ATE as the parameter of interest:

\begin{equation*}
  \Psi_0 = \E_0(Y_1) - \E_0(Y_0).
\end{equation*}

This parameter of interest is identifiable under following assumptions:
\begin{assumption}[Consistency]
  \label{assum:consistent}
  \begin{equation*}
    Y_i = Y_{i,A_i} = Y_{i0}(1-A_i) + Y_{i1}A_i.
    \end{equation*}
\end{assumption}
\begin{assumption}[Conditional Randomization]
  \label{assum:cr}
  \begin{equation*}
    (Y_0, Y_1) \indep A | W.
    \end{equation*}
\end{assumption}
Assumption \ref{assum:cr} has also been called strong ignorability, or unconfoundedness \citep{rubin1990formal}. Under assumption \ref{assum:consistent} and \ref{assum:cr}, the conditional probability of $Y = y$ given $A = a, W = w$ can be written as:

\begin{equation*}
P(Y = y|A = a, W = w) = P(Y_a = y|W = w),
  \end{equation*}
thus the conditional expectation of $Y$ given $A = a, W  = w$ can be written as:
\begin{equation*}
\E(Y|A = a, W = w) = \E(Y_a|W = w),
  \end{equation*}
and the parameter of interest, ATE, can be written as:
\begin{equation*}
  \begin{aligned}
    \Psi_0 &= \E(Y_1) - \E(Y_0) \\&= \E_0(\E_0(Y|A = 1,W)) -\E_0(\E_0(Y|A = 0,W))
    \end{aligned}
\end{equation*}

\begin{assumption}[Positivity]
  \label{assum:pos}
  \begin{equation*}
   0 < g_0(W) < 1
  \end{equation*}
  almost everywhere.
\end{assumption}
Assumption \ref{assum:pos} is necessary for the identification. Otherwise, the model is not identifiable, as we can never observe one of the potential outcomes for the units with certain baseline covariates $W$.

\section{Brief Review of Some Common Estimators}
\label{sec:review}

One of the well-studied estimators for the ATE in observational studies is the G-computation estimator (or outcome regression model), which estimates $\bar{Q}_0$ with $\bar{Q}_n$, and then estimates ATE by the following formula:

\begin{equation*}
  \Psi_n^{G-comp} = \frac{1}{n}\sum_{i=1}^{n} \left[\bar{Q}_n(A = 1, W = W_i) - \bar{Q}_n(A = 0, W = W_i)\right].
\end{equation*}
As long as the aforementioned assumptions hold, and the conditional response model estimator $\bar{Q}_n$ for $\bar{Q}_0$ is consistent, the resulting estimator $\Psi_n^{G-comp}$  is also consistent.

Another widely used estimator is the Inverse Probability of Treatment Weighting (IPW) estimator. It only relies on the estimator $g_n$ of $g_0$:

\begin{equation*}
  \begin{aligned}
    \Psi_n^{IPW} = \frac{1}{n}\sum_{i=1}^{n} \left[ \frac{A_iY_i}{g_n(W_i)} - \frac{(1-A_i)Y_i}{1- g_n(W_i)} \right] ,
    \end{aligned}
\end{equation*}
where $g_n$ is usually fitted by a supervised model (e.g. logistic regression), which regresses $A$ on the pre-treatment confounders $W$. Similar to G-computation, the IPW estimator is consistent as long as all of the aforementioned assumptions hold, and the estimated PS,  $g_n$, is consistent. However, the IPW estimator can be highly unstable since extreme values of the estimated PS can lead to overly large and unstable weights for some units. This phenomenon is called the practical positivity violation. To overcome this issue, \citep{hajek1971comment} proposed a stabilized estimator:

\begin{equation*}
  \begin{aligned}
    \Psi_n^{Hajek-IPW} = \sum_{i=1}^{n} \left[  \frac{A_iY_i/g_n(W_i)}{\sum_{i=1}^{n} A_i/g_n(W_i)} \right.
      \\
     \left.  - \frac{(1-A_i)Y_i/(1- g_n(W_i))}{\sum_{i=1}^{n}(1-A_i)/(1- g_n(W_i))} \right],
    \end{aligned}
\end{equation*}
where the denominator $n$ is replaced by the weight normalization term $A_ig_n(W_i)$ and $(1-A_i)(1-g_n(W_i))$. It is easy to show that this estimator is also consistent as long as $g_n$ is a consistent estimator.

All of the estimators mentioned above are not robust in the sense that misspecification of the first stage modeling (of conditional outcome, or the PS) could lead to biased estimation for the causal parameter of interest. This is the reason why double robust (DR) estimators are preferable. DR estimators  usually rely on the estimation of both $\bar{Q}_0$ and $g_0$. As long as one of them is estimated consistently, the resulting final estimator would be consistent. Weighted Regression (WR) is one of the commonly used DR-estimators \citep{kang2007demystifying,bang2005doubly}. In comparison to G-computation, it estimates $\bar{Q}_0$ by minimizing the weighted empirical loss:

\begin{equation*}
  \bar{Q}_n^{WR} = \argmin_{\bar{Q}}\sum_{i=1}^{n}[\omega_{i}(g_n)L(\bar{Q}(A_i,W_i), Y_i)]
  \end{equation*}
where the weight is defined $\omega_{i}(g_n) = [A_i/g_n(W_i)+ (1-A_i)/(1-g_n(W_i))]$, and $L$ is the loss function. The estimator for the causal parameter is defined as:
\begin{equation*}
  \begin{aligned}
    \Psi_n^{WR} = \frac{1}{n}\sum_{i=1}^{n} [\bar{Q}_n^{WR}(A = 1, W = W_i)\\
      - \bar{Q}_n^{WR}(A = 0, W = W_i)].
    \end{aligned}
\end{equation*}
The WR estimator is also called the weighted least squares (WSL) estimator, if the loss function is the squared error $L(x,y) = (x-y)^2$.

Augmented IPW (A-IPW, or DR-IPW) is another DR-estimator which can be written as:

\begin{equation}
  \label{aipw}
\begin{aligned}
  \Psi_n^{DR-IPW} = \frac{1}{n}\sum_{i=1}^{n} H_{g_n}(A_i,W_i)[Y_i - \bar{Q}_n(A_i,W_i)]
  \\+ \bar{Q}_n(1,W_i) - \bar{Q}_n(0,W_i)
\end{aligned}
\end{equation}
where
\begin{equation*}
H_{g_n}(A,W) = \frac{A}{g_n(W)} - \frac{1-A}{1 - g_n(W)},
\end{equation*}

\trc{is designed based on the target parameter, ATE.} $\Psi_n^{DR-IPW}$ also relies on both $\bar{Q}_n$ and $g_n$.  It was first proposed by \citep{cassel1976some,cassel1977foundations} where it was called the ``bias-corrected estimator''. It corrects the bias from the initial estimate, $\bar{Q}_n $, with the weighted residual 
from the initial fit. \citep{robins1994estimation}  proposed a class of estimators which contains \ref{aipw}, and \citep{robins1995semiparametric} further showed that \ref{aipw} is a locally semiparametric efficient estimator.

Similar to the IPW estimator, A-IPW is also influenced by extreme weights, as it uses inverse probability weighting. However, a Hajek style stabilization could mitigate concerns of overly influential weights:

\begin{equation}
  \begin{aligned}
  \label{bch}
  \Psi_n^{HBC-IPW} = \sum_{i=1}^{n} \left[ \left(\frac{A/g_n(W)}{\sum_{i=1}^{n} A/g_n(W)}\right. \right.\\
    \left. -  \frac{(1-A)/(1 - g_n(W))}{\sum_{i=1}^{n}(1-A)/(1 - g_n(W))} \right) (Y_i -\bar{Q}_n(A_i,W_i)
 \\ \left. + \frac{1}{n} \bar{Q}_n(1,W_i) - \frac{1}{n} \bar{Q}_n(0,W_i) \right].
  \end{aligned}
  \end{equation}
For simplicity, we will call the estimator in equation \ref{bch} the HBC (Hajek type bias-correction) estimator. Although this estimator no longer enjoys some attractive theoretical properties (e.g. efficiency) of A-IPW, it is still DR, and it can potentially improve  finite sample performance. It is also possible that these modifications could improve the robustness of the estimated treatment effects when both of the models are misspecified \citep{kang2007demystifying}.

\section{Brief Review of Targeted Minimum Loss-based Estimation (TMLE)}
\label{sec:tmle}

Targeted minimum loss-based estimation is a general template to estimate a user-specified parameter of interest, given a user-specified loss function, and fluctuation sub-model. In this study, we consider the ATE as our target parameter, the negative likelihood as the  loss function, and the logistic fluctuation. Let $Y$ represent a binary variable, or a continuous variable within the range $(0,1)$ \footnote{otherwise, we could simply normalize $Y$ into $(0,1)$ and finally rescale the  estimate $\Psi_n^{TMLE}$ back }. The TMLE estimator for the ATE can be written as:

\begin{equation}
  \label{eq:tmle}
\Psi_n^{TMLE} = \frac{1}{n}\sum_{i=1}^{n} (\bar{Q}_n^*(1, W_i) - \bar{Q}_n^*(0, W_i)).
\end{equation}

 In equation \ref{eq:tmle}, $\bar{Q}_n^*$ (which is within the range $(0,1)$) is updated from an initial estimate, $Q_n$, by a logistic fluctuation sub-model:
 \begin{equation}
   \label{eq:target}
\logit(\bar{Q}_n^*(A,W)) = \logit(\bar{Q}_n(A,W)) + \epsilon H_{g_n}(A,W).
\end{equation}
The fluctuation parameter $\epsilon$ is estimated through maximum likelihood estimation, or equivalently, minimizing the negative log-likelihood loss:

\begin{equation*}
L(\epsilon) = \sum_{i=1}^{n}  Y_i(\bar{Q}_n^*(A_i, W_i)) + (1 - Y_i) (1 - \bar{Q}_n^*(A_i, W_i)).
\end{equation*}
If either the propensity model or outcome model is consistent, then the TMLE estimator is consistent. If both of them are consistent, then the TMLE estimator is also efficient. To consistently estimate $\bar{Q}_0$ and $g_0$, we suggest  using Super Learner, a  data-adaptive ensemble method, for prediction modeling \citep{van2007super,polley2010super,pirracchio2015improving,ju2016propensity,benkeser2016online,ju2017relative}.

In addition to double robustness and asymptotic efficiency, TMLE has following advantages:

\begin{enumerate}
\item Equation \ref{eq:tmle} shows that TMLE is a plug-in estimator and therefore, respects the global constraints of the model. For instance, suppose $Y$ is binary. The ATE, therefore, should be between $[-1,1]$. However, some competing estimators may produce estimates out of such bounds. Since TMLE maps the targeted estimate $P^*$ of $P_0$ into the mapping $\Psi$, it respects knowledge of the model.

\item The targeting step in TMLE is a minimum loss estimation \footnote{it is maximum likelihood estimation (MLE) if the loss is negative log-likelihood}, which offers a metric to evaluate the goodness-of-fit for $g_n$ and $\bar{Q}_n$, w.r.t. the parameter of interest $\Psi_0$. 

  \item In the empirical/simulation studies by \citep{porter2011relative}, \trc{TMLE is more robust than IPW and A-IPW to positivity, or near positivity, violations , where $g_n$ is too close to 0 or 1.}
  
  \end{enumerate}

\section{Brief Review of Collaborative TMLE}
\label{sec:ctmle}

\subsection{C-TMLE for Variable Selection}

In the TMLE algorithm, the estimate of $\bar{Q}_0$ is updated by the fluctuation step, while the estimate of $g_0$ is estimated externally and then held fixed. One  extension of TMLE is to find a way to estimate $g_{0}$ in a \emph{collaborative} manner. Motivated by the second advantage of TMLE, collaborative TMLE  was proposed to make this extension feasible \citep{van2010collaborative}. Here we first briefly review the general template for C-TMLE:

\begin{enumerate}
\label{template}
\item Compute the initial estimate $\bar{Q}_n^0$ of $\bar{Q}_0$.
\item \trc{Compute a sequence of estimates $g_{n, k}$ and $\bar{Q}_{n,k}^{*}$ for $g_0$ and $\bar{Q}_0$ respectively, with $k = 1, \ldots, K$. With $k$ increasing, the empirical loss for both $g_{n,k}$ and $\bar{Q}_{n,k}^{*}$ would decrease. In addition, we require $g_{n,K}$ to be asymptotically consistent for $g_{0}$.}
\item  Build a sequence of TMLE candidate estimators, based on a given fluctuation model.

\item Use cross-validation for  step 3 to select the  $\bar{Q}_{n,k}^{*}$, that minimizes the cross-validated risk, and denote this TMLE estimator as the C-TMLE estimator.
  
\end{enumerate}

This is a high-level template  for the general C-TMLE algorithm. There are many variations of instantiations of this template. For example the greedy C-TMLE was proposed by \citep{van2010collaborative,gruber2010application} for variable selection in a discrete setting. The following are some details of greedy C-TMLE:

\begin{itemize}
\item In step 2, the greedy C-TMLE algorithm starts from an intercept model (which fits the PS with its mean), and then builds the sequence of $g_{n,k}$ by using a forward selection algorithm: during each iteration $k$, for each of the remaining covariates $W_j$, that have not been selected yet, we add it into the previous PS model $g_{n, k-1}$, which yields a larger PS model $g_{n,k}^j$ and $H_{g_{n,k}^j}$. We then compute  $\bar{Q}_{n,k}^{*,j}$ by equation \ref{eq:target}. For all $j$, we select the PS model that corresponds to the $\bar{Q}^{*,j}_{n,k}$ with the smallest empirical loss. For simplicity we call this the forward selection step at the $k$-th iteration.

  \item \label{itm:target} For the initial estimate in equation \ref{eq:target}, we start with $\bar{Q}_{n,1} = \bar{Q}_n$. For each iteration $k$, we first try $\bar{Q}_{n,k} = \bar{Q}_{n,k-1}$. If all of the possible  $\bar{Q}_{n,k}^{*,j}$ mentioned above do not improve the empirical fit compared to $\bar{Q}_{n,k-1}^{*}$, we update $\bar{Q}_{n,k} = \bar{Q}_{n,k-1}^{*}$ and rerun the forward selection step at the $k$-th iteration. \trc{Notice, that as we use the last TMLE estimator as the candidate, all of the current candidate $\bar{Q}_{n,k}^{*,j}$ are guaranteed to have a better empirical fit compared to their initial estimate $\bar{Q}_{n,k}$. Otherwise if there is at least one candidate that improves the empirical fit, we just move to the next forward selection step.} In this manner, we make sure that the empirical loss for each candidate $\bar{Q}_{n,k}^*$ is monotonically decreasing.
\end{itemize}
          
\citep{ju2016scalable} also proposed scalable versions of the discrete C-TMLE algorithm as new instantiations of the C-TMLE template. These scalable C-TMLE algorithms  avoid the forward selection step by enforcing a user-specified ordering of the covariates. \citep{ju2016scalable} showed that these scalable C-TMLE algorithms have all of the asymptotic theoretical properties of the greedy C-TMLE algorithm, but with much lower time complexity.

\subsection{C-TMLE for  Model Selection of LASSO}

To the best of our knowledge, C-TMLE has primarily been applied for variable selection. However, it can easily be adapted to more general model selection problems. In our recent work \citep{ju2017ctmle,van2017ctmle}, two instantiations of the C-TMLE algorithm were proposed for a general model selection problem with a one-dimensional hyper-parameter. In this study, we consider an example where the PS model is estimated by LASSO:

\begin{equation*}
  \begin{aligned}
  \beta_{n,\lambda} &= \min _{\beta \in \mathbb {R}^{p}}\left(\frac{1}{n}\sum_{i=1}^{n} L(A_i, \logit( \beta W_i))+\lambda \|\beta \|_{1}\right)
\\ g_{n, \lambda}(W_i) &= \logit(\beta_{n,\lambda} W_i)
  \end{aligned}
\end{equation*}

where $L$ is the negative log-likelihood for the Bernoulli distribution, as $A$ is binary. We used C-TMLE to select the PS estimator, $g_{n,\lambda}$, with the best penalty parameter $\lambda$. We applied two C-TMLE algorithms for model selection of LASSO.  Here, we provide a brief outline for each of the algorithms. Details are provided in the supplemental appendices.

\begin{itemize}

\item C-TMLE1: First, we briefly introduce the C-TMLE1 algorithm. According to the C-TMLE template outlined above, C-TMLE1 first builds an initial estimate for $\bar{Q}_n$ and a sequence of  propensity score estimators, $g_{n,\lambda_k}$, for $k \in 0, \ldots, K$, each with a penalty $\lambda_k$, where $\lambda_k$ is monotonically decreasing. \trc{We recommend to set $\lambda_{1} = \lambda_{CV}$ because the cross-validation usually selects the over-smoothed PS estimator, thus it is unnecessary to consider $\lambda_{1} >\lambda_{CV}$.} Then, we just follow  step 3 in the template  described previously, and build a sequence  of estimators, $\bar{Q}_{n,\lambda}^*$, each corresponding to $g_{n,\lambda}$. We then select the best $\bar{Q}^*_{n,\lambda_{ctmle}}$ by using cross-validation, with its corresponding initial estimate $\bar{Q}_{n,\lambda_{ctmle}}$. Finally we fluctuate the selected initial estimate $\bar{Q}_{n,\lambda_{ctmle}}$ with  each $g_{n,\lambda}$  for $ \lambda_K < \lambda< \lambda_{ctmle}$, yielding a new sequence $\bar{Q}_{n,\lambda}^{*}$. We choose  $\bar{Q}_n^* = \bar{Q}_{n,\lambda}^{*}$ ,  which minimizes the empirical loss, as our final estimate.  The final step guarantees that a critical equation:
  \begin{equation}
    \begin{aligned}
  \label{eq:critical}
  &P_n D^+(\bar{Q}_{n,\lambda}^*,g_{n,\lambda})\\ =& \frac{\partial }{\partial \lambda}\sum_{i=1}^{n} H_{g_{n,\lambda}}(A_i,W_i)(Y_i - \bar{Q}_{n,\lambda}^{*}(A_i, Y_i))=0
  \end{aligned}
  \end{equation}
  is solved \citep{ju2017ctmle,van2017ctmle}. \trc{This guarantees that the resulting C-TMLE estimator is asymptotically linear under regularity conditions even when $\bar{Q}_n$ is not consistent.} A detailed description of C-TMLE1 is provided  in appendix \ref{sec:ctmle1}.

\item C-TMLE0: the C-TMLE0 algorithm does not select the PS estimator collaboratively. Instead, it is exactly the same as the TMLE algorithm, except it updates the estimate by equation \ref{eq:ctmle}:
  \begin{equation}
    \begin{aligned}
   \label{eq:ctmle}
   \logit(\bar{Q}_n^*(A,W)) = \logit(\bar{Q}_n(A,W)) \\+ \epsilon_1 H_{g_{n,\lambda_k}}(A,W) + \epsilon_2 \tilde{H}_{g_{n,\lambda_k}}(A,W)
   \end{aligned}
 \end{equation}
 where
 \begin{equation*}
   \begin{aligned}
    \tilde{H}_{g_{n,\lambda_k}}(A,W) &= \frac{\partial H_{g_{n,\lambda}}(A,W)}{\partial \lambda}\at{\lambda = \lambda_k} \\
    &=
    \frac{1-A}{(1 -g_{n,\lambda_k}(W))^2}\frac{\partial ( 1 - g_{n,\lambda})}{\partial \lambda}\at{\lambda = \lambda_k} \\&+
    \frac{A}{g_{n,\lambda}(W)^2} \frac{\partial g_{n,\lambda_k}}{\partial \lambda}\at{\lambda = \lambda_k}.
    \end{aligned}
 \end{equation*}

 Note we still call it C-TMLE as it solves the critical equation \ref{eq:ctmle}.  Solving the additional clever covariate $\tilde{H}_{g_{n,\lambda_k}}(A,W)$ could be considered as an approximation of the collaborative selection in C-TMLE1 \citep{ju2017ctmle,van2017ctmle}. More details of C-TMLE0 can be found in appendix \ref{sec:ctmle3}.
  
\end{itemize}

\section{Data Source}
\label{sec:data}

In previous work by \citep{ju2016propensity}, Super Learner was applied to three electronic healthcare data sets for propensity score estimation. In two of the data sets (NOAC  study and Vytorin study), the PS model showed strong non-linearity patterns, where non-linear algorithms (gbm) outperformed main term LASSO (w.r.t. the predictive performance of the estimated PS) with the same covariate set. Thus the main term linear model may result in strong model misspecification for such a dataset.  To better demonstrate C-TMLE for LASSO selection under mild model misspecification, we only considered the NSAID dataset, where the treatment mechanism could be estimated satisfactorily with main term linear models. \trc{This data set  was first created by  \citep{brookhart2006evaluating}, and further studied by \citep{schneeweiss2009high,rassen2012using}.}

\subsection{Nonsteroidal Anti-inflammatory Drugs Study}

In this study, the observations were sampled from a population of patients aged 65 years and older who were enrolled in both Medicare and the Pennsylvania Pharmaceutical Assistance Contract for the Elderly (PACE) programs between 1995 and 2002. The treatment is a binary indicator  taking on values of 1 for patients who received a selective COX-2 inhibitor and 0 for patients who received a non-selective nonsteroidal anti-inflammatory drug. The outcome is also a binary indicator  taking on values of 1 for patients who are diagnosed with gastrointestinal (GI) complications during the follow-up periods, and 0 otherwise.

\trc{To adjust for potential confounders}, some predefined baseline pre-treatment covariates were collected (e.g. age, gender, race). To further adjust for confounding we implemented a widely used variable selection algorithm for healthcare claims databases, known as the high-dimensional propensity score (hdPS) (discussed further below) \citep{schneeweiss2009high}. The dataset for this study included $9,470$ claims codes, which were clustered into 8 categories, including ambulatory  diagnoses, ambulatory procedures,  hospital diagnoses, hospital procedures,  nursing home  diagnoses, physician  diagnoses, physician procedures  and  prescription drugs. The value for each claims code denotes the number of times the respective patient received the healthcare procedure corresponding to the code during a 12 month baseline period prior to treatment initiation. Thus all of the claims data are non-negative integers.

\begin{table}[H]
  \centering
  \caption{Brief summary of the NSAID study databases}
  \label{table:nsaid}
  \begin{tabular}{|l|l|}
    \hline
    Sample Size  & $49,653$    \\ \hline
    \# of  Baseline Covariates & 22        \\ \hline
    \# of Code Resource  & $8$       \\ \hline
    \# of Claims Code  & $9,470$       \\ \hline
  \end{tabular}
\end{table}

\subsection{The High-Dimensional Propensity Score (hdPS) to Learn from Health Insurance Data}
\label{subsec:hdps}

Claims data are usually high-dimensional ($p_c = 9,470$ in this study) due to large amounts of healthcare diagnoses and procedures. Further, claims data are often  highly sparse as each patient often receives only a few diagnoses. To address these issues, the hdPS variable selection algorithm was introduced by \citep{schneeweiss2009high}  to generate hundreds of baseline variables from claims codes, and then rank them by their potential confounding impact. Its core part is outlined in the  following steps:

\begin{enumerate}
\item Cluster the codes according to  their source \footnote{We  replace the term ``data dimension'' in \citep{schneeweiss2009high} with ``source'' to avoid ambiguity.}: this is determined manually based on the origin and quality of data feeds and is unique to the database being used. In this study, the codes come from 8 sources.
  
\item Identify candidate codes in each cluster: for each code count $c$, compute its empirical prevalence $p_{n,c} = \E_nI(c >0)$, rank all covariates by $\max(p_{n,c}, 1 - p_{n,c})$, and select the top $k_1$ codes within each cluster. In the NSAID study, we have $8k_1$ claims covariates left after this step.

\item Generate hdPS covariates: For each claims covariates, $c_i$, for each individual, $i$, construct three indicator variables where:  $c_{i}^{(1)}$ is equal to  one if and only if (iff) $c_{i}$ is positive, $c_{i}^{(2)}$ is equal to one iff   $c_{i}$  is  larger  than the  median  of  $\{c_{i}  :  1\leq i  \leq  n\}$, and $c_{i}^{(3)}$ is equal to  one iff $c_{i}$ is larger than  the 75\%-quantile of $\{c_{i}  : 1\leq  i \leq  n\}$. We denote these new covariates as ``hdPS covariates''. For the empirical example in this study, this step results in $24k_1$ generated hdPS covariates.

\item Select hdPS covariates  for confounding adjustment: Use the Bross formula \citep{bross54,schneeweiss2009high} to rank each hdPS covariate, $c$, by its potential for confounding bias:

    \begin{equation*}
    \text{Bias}(c)=\frac{\E_n(c=1|A = 1) (rr_{n}(c) - 1) + 1}{\E_n(c=1|A = 0) (rr_{n}(c) - 1) + 1}
  \end{equation*}
  with
  \begin{eqnarray*}
   rr_n(c)
    &=& 
   \frac{\E_n(Y = 1|c = 1)}{\E_n(Y = 1|c = 0)}
  \end{eqnarray*}

where  $\E_n$ denotes the empirical distribution of data.

Covariates are then ranked by descending order of $|\log(\text{Bias}(c))|$. We then select the first $k_2$ ordered hdPS covariates among   the total $24k_1$ hdPS (generated) covariates from  step 3.
\end{enumerate}

\trc{
The hdPS algorithm has been used in studies evaluating the effectiveness of prescription drugs and medical procedures using healthcare claims data in the U.S. \citep{schneeweiss2010comparative,patorno2014studies,le2013effects,kumamaru2016comparison}, Canada \citep{filion2013proton,dormuth2014higher,guertin2016performance} , Europe \citep{garbe2013high,hallas2017performance,enders2017potential},  and electronic health records \citep{neugebauer2015high, toh2011confounding}.  \citep{schneeweiss2017variable} evaluated a range of algorithms to improve covariate ranking based on the empirical covariate outcome relationship without any meaningful improvement over the ranking using the Bross formula. \citep{ju2016propensity} evaluated various choices for the parameters $k_1$ and $k_2$ within the hdPS algorithm, and found that the performance of the hdPS was not sensitive to choices for $k_1$ and $k_2$ as long as the hyper-parameter pair were within a reasonable range. For this study, we let $k_1=100$ and $ k_2 = 200$. For simplicity, we denote the combined set of predefined baseline covariates and selected hdPS covariates as $W$.
}


\section{Quasi-Experiment}
\label{sec:quasi}

\subsection{Simulation Setting}

In this simulation, we generated partially synthetic data based on the NSAID data set. We designed our own conditional distribution of the outcome, $Y$, given treatment, $A$, and baseline covariates, $W$, while keeping the structure of the treatment mechanism $g_0(A|W)$ so that the relationships between covariates with treatment assignment were preserved \citep{franklin2014plasmode}. In our study, the conditional distribution of the outcome was defined as:

\begin{equation}
  \label{eq:Q}
  Y_i =  2 + \beta W_i + A_i + \epsilon_i
\end{equation}
where $\epsilon_i$ is drawn independently from the standard normal distribution. We then selected 40 covariates that had the highest Pearson correlation with treatment $A$. The coefficient of $\beta$ in equation \ref{eq:Q} was set to zero for all the non-selected covariates. The coefficient for the selected covariates was sampled from separate and independent standard normal distributions, and were fixed across all simulations. We define the marginal distribution of $W$ as the empirical distribution of $W_i$ for $i \in 1 \ldots n$. The parameter of interest is the ATE, thus it is identifiable if we know the distribution of the conditional response $Y|A,W$ and marginal distribution of $W$.

In our simulation, we considered two settings. In the first setting, only the first 10 out of 40 confounders were used to estimate $\bar{Q}_0$. In the second setting, $\bar{Q}_0$ was estimated using the first 20 out of 40 confounders.

By the description above, we  have the following:

\begin{itemize}
  
\item There are only 40 confounders in total. 

\item The true value of the parameter of interest (ATE) is 1.

\item The treatment mechanism $g_0(A|W)$ comes from a real world data generating distribution, which is usually non-linear.  \citep{ju2016propensity} showed that the PS in this example can be estimated well by linear models. Therefore, in this example the PS model is only mildly misspecified.

\item Both $\bar{Q}_0$ and $g_{0}$ are estimated with a misspecified model: $\bar{Q}_0$ is estimated with an incomplete predictor set; $g_{0}$ is estimated with linear model, while there is no reason to believe it is truly linear. 
  
\end{itemize}

The results are computed across 500 replications, each with sample sizes of 1000.

\subsection{Competing Estimators}

In this study, we focused on PS based estimators, including inverse probability of treatment weight (IPW) estimator, Hajek type IPW estimator, double robust (augmented) inverse probability of treatment weight (DR-IPW, or A-IPW) estimator, Hajek type Bias-correction  (HBC) Estimator, weighted regression (WR) estimator, targeted maximum likelihood estimator (TMLE), and the proposed two collaborative-TMLE estimators.

For all PS based estimators, we consider two variations. For the first variation, we first used the cross-validated LASSO  (CV-LASSO) algorithm to find the regularization parameter $\lambda_{CV}$ of LASSO for PS estimation, and then plugged it into the final estimators. In the second variation, we first applied C-TMLE1, and use LASSO with the regularization parameter $\lambda_{C-TMLE}$ selected by C-TMLE1 to estimate the PS, and then plug it into the estimator. Taking IPW as example, we used ``IPW'' to denote the first variation, and ``IPW*'' for the second variation.

It is important to note that in  this case, ``TMLE*'' is actually a variation of collaborative TMLE, as the PS model is selected collaboratively \citep{gruber2010application,van2010collaborative}. However, it is different from the proposed C-TMLE algorithms, as it does not solve the critical equation \ref{eq:critical}.

It is also important to note that both C-TMLE and CV-LASSO use cross-validation. For simplicity, and to avoid ambiguity, we use term ``CV'' to denote the non-collaborative model selection procedure which relies on the cross-validation w.r.t. the prediction performance for the treatment mechanism itself (e.g. the model selection step in CV-LASSO).

\subsection{Point Estimation}
\label{subsec:point_est}

We first compared the variance, bias, and mean square error (MSE) for the point estimation from all the competing estimators in two settings.

\begin{figure}[ht]
  \centering
  \begin{subfigure}[b]{0.35\textwidth}
    \includegraphics[width=\textwidth]{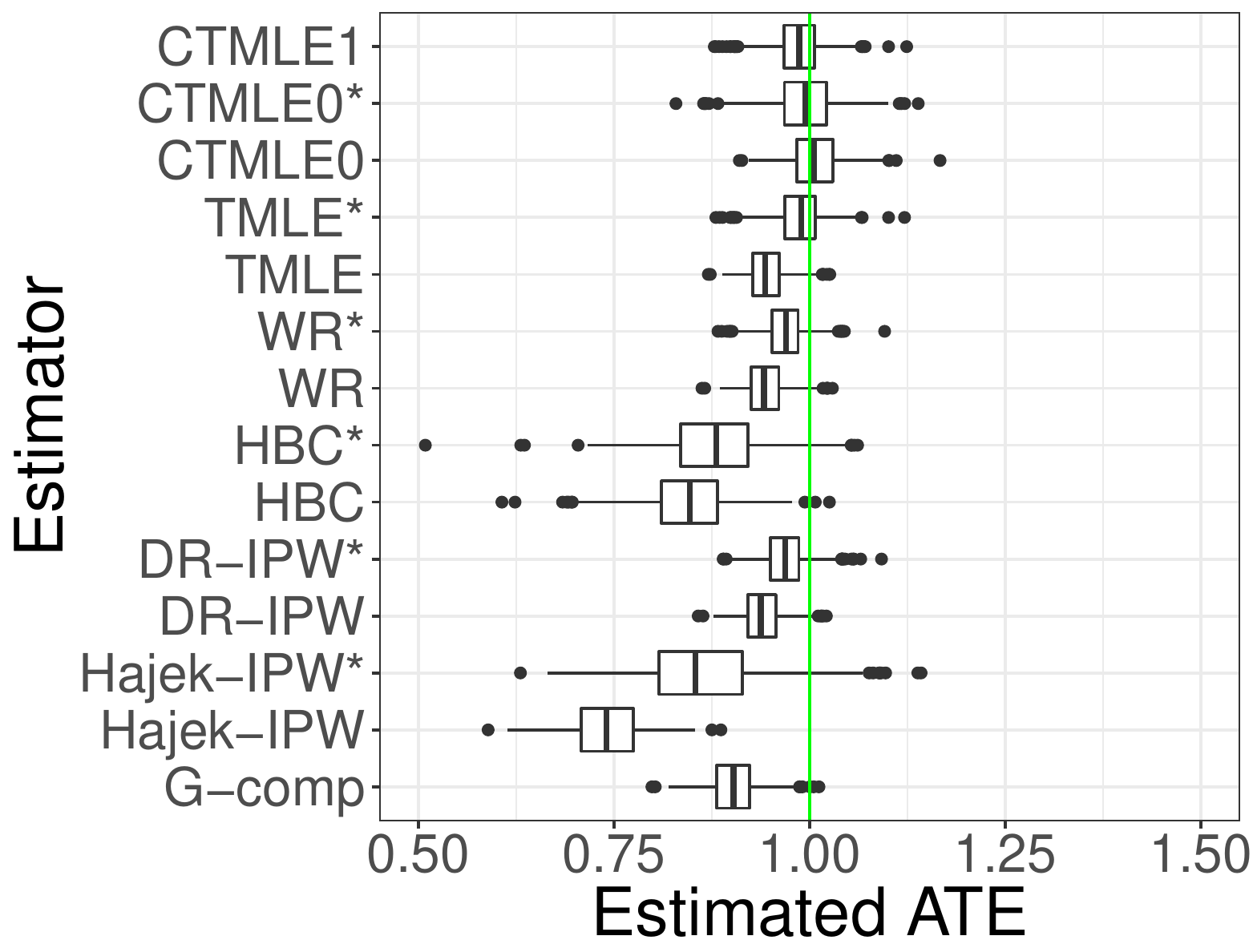}
  \end{subfigure}
  \hspace{2mm}
  \begin{subfigure}[b]{0.35\textwidth}
    \includegraphics[width=\textwidth]{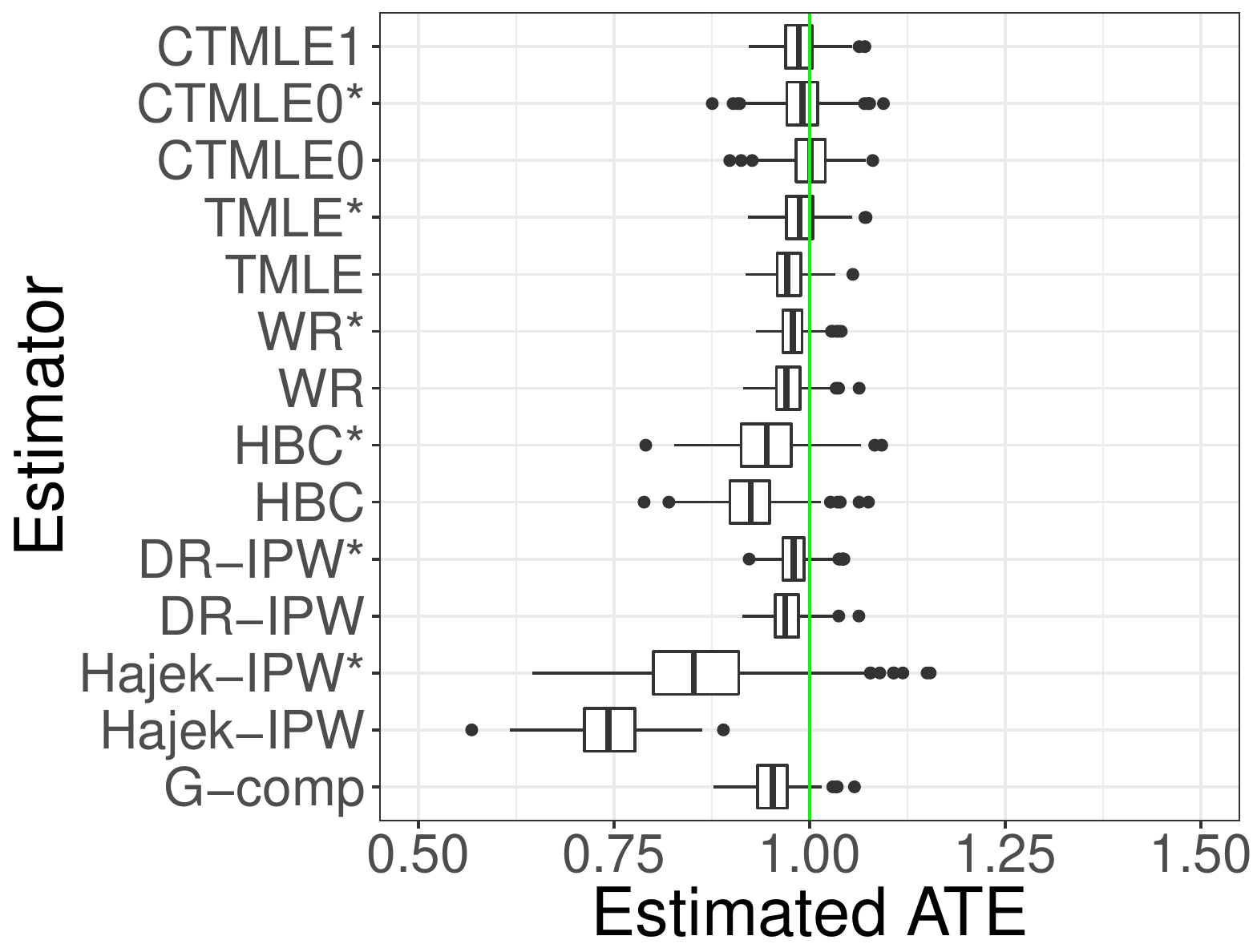}
  \end{subfigure}
  \caption{Boxplot of the estimated ATE for each estimator across 500 replications, when the initial estimate is fit on 10/20 out of 40 confounders.}
  \label{fig:sim_all}
\end{figure}

\begin{table}[ht]
  \centering
  \caption{Performance of Point Estimation for Estimators when the initial estimate $\bar{Q}_n$ of $\bar{Q}_0$ is estimated  on 10 and 20 out of 40 confounders. The results are computed based on simulations across 500 replications, each with a sample size of 1000 based on the NSAID study. All of the numeric values are on a scale of $10^{-2}$. }
   \label{tab:nsaid_est}
  \scalebox{0.8}{
  \begin{tabular}{|r|r|rrrrrr|}
    \hline
    Initial Fit& & unadj & G-comp & WR & WR* & Hajek-BC & Hajek-BC*  \\
    \hline
    10/40 &Bias & -59.29 & -9.69 & -5.68 & -3.11 & -15.54 & -12.29 \\ 
    & SE & 8.43 & 3.36 & 2.66 & 2.75 & 5.80 & 6.63 \\ 
    & MSE & 35.87 & 1.05 & 0.39 & 0.17 & 2.75 & 1.95 \\
    \hline
    20/40 & Bias & -59.91 & -4.72 & -2.77 & -2.12 & -7.56 & -5.47 \\ 
    & SE & 8.36 & 2.73 & 2.27 & 1.92 & 4.10 & 4.54 \\ 
    & MSE & 36.59 & 0.30 & 0.13 & 0.08 & 0.74 & 0.51 \\ 
    \hline
    Initial Fit& & IPW & IPW* & Hajek-IPW & Hajek-IPW* & DR-IPW & DR-IPW* \\
    \hline 
    10/40&  Bias & 95.43 & 128.97 & -25.86 & -13.61 & -6.07 & -3.12 \\ 
    & SE & 36.55 & 91.38 & 4.85 & 8.21 & 2.63 & 3.02 \\ 
    & MSE &  104.40 & 249.69 & 6.92 & 2.53 & 0.44 & 0.19  \\ 
    \hline
    20/40 &  Bias & 97.11 & 125.85 & -25.60 & -13.70 & -2.92 & -1.95 \\ 
    &SE & 35.98 & 90.85 & 4.77 & 8.56 & 2.26 & 2.17 \\ 
    & MSE & 107.23 & 240.75 & 6.78 & 2.61 & 0.14 & 0.09  \\ 
    \hline
    Initial Fit & & TMLE & TMLE* & CTMLE1 & CTMLE0 & CTMLE0*& \\
    \hline
    10/40 &  Bias & -5.49 & -1.23 & -1.40 & 0.70 & -0.64& \\
    & SE & 2.57 & 3.46 & 3.56  & 3.38 & 4.40 &\\ 
    & MSE & 0.37 & 0.13 & 0.15  & 0.12 & 0.20&  \\
    \hline
20/40 &Bias & -2.68 & -1.28 & -1.38 & 0.08 & -0.95 &\\ 
 & SE & 2.19 & 2.53 & 2.53  & 2.85 & 3.07 &\\ 
 & MSE & 0.12 & 0.08 & 0.08 & 0.08 & 0.10 &\\ 
    \hline
  \end{tabular}
 }
\end{table}

Table \ref{tab:nsaid_est} and figure \ref{fig:sim_all} show the performance of all the competing estimators. IPW has very large variance and bias, which might due to the violation of the positivity assumption. We can see that  TMLE*, C-TMLE1, CTMLE0, and CTMLE0* outperformed other estimators, with each having similar performance. In addition, C-TMLE0* did not show any improvement compared to C-TMLE0. This is consistent with previous results \citep{ju2017ctmle,van2017ctmle}.

We also evaluated  the relative performance of other PS based estimators with $g_n$  selected by C-TMLE, compared with $g_n$ selected by CV. For IPW, the performance was still poor. However, for all of the other estimators that rely on the estimated PS, the performance improved considerably. Taking the first setting as an example, the relative empirical efficiency of DR-IPW* compared to DR-IPW was $\frac{\text{MSE(DR-IPW)}}{\text{MSE(DR-IPW*)}} = 1.52$, while for TMLE it was $\frac{\text{MSE(TMLE)}}{\text{MSE(TMLE*)}} = 1.66$. The relative empirical efficiency for both of these estimators is improved with a reduction in bias and slight increase in variance. These empirical results are consistent with previous theory \citep{ju2017ctmle,van2017ctmle} showing that the model selected by external CV is usually \trc{over-smoothed}. These results illustrate the weakness of using ``external'' CV for PS model selection. 


\subsection{Confidence Interval}
\label{subsec:ci}

In this section, we evaluate the coverage and the length of the confidence intervals (CIs) for all the double robust  estimators.

\begin{table}[ht]
    \caption{Coverage of the 95\% confidence intervals for semi-parametric efficient estimators when the initial estimate $\bar{Q}_n$ of $\bar{Q}_0$ is estimated  on 10 and 20 out of 40 confounders. The results are computed across 500 replications, each with sample sizes of 1000 based on the NSAID study. All of the numerical values are multiplied by 100.}
  \label{tab:nsaid_ci}
  \centering
  \scalebox{0.85}{
    \begin{tabular}{|r|r|rrrrrrr|}
      \hline
      &  & CTMLE1 & CTMLE0 & CTMLE0* & DR-IPW & DR-IPW* & TMLE & TMLE*\\ 
      \hline
      10/40& Coverage  & 0.926  & 0.920 & 0.910 & 0.458 & 0.914  &   0.526 & 0.942\\
      & Average Length  & 0.142  & 0.115 & 0.142 & 0.120 & 0.159 &  0.119 & 0.144 \\
      \hline
      20/40 & Coverage & 0.934    & 0.872 & 0.898 & 0.748 & 0.928  & 0.790 & 0.946 \\
      & Average Length & 0.105  & 0.087 & 0.103 & 0.088 & 0.112 & 0.087 & 0.106\\  
      \hline
    \end{tabular}
  }
\end{table}

In both settings,  TMLE* and C-TMLE1 had the best coverage. We can see that for other estimators, the length of the CIs were usually smaller/under-estimated. This resulted in a  less satisfactory coverage even though the point estimation had similar performance (e.g. compare C-TMLE0 to C-TMLE1). With collaboratively selected $g_n$, the coverage of TMLE and DR-IPW  improved significantly. These empirical results illustrate that a more targeted propensity score model selection can improve both causal estimation and inference.

\subsection{Pairwise Comparison of Efficient Estimators}
\label{subsec:pair}

In this subsection, we studied the pairwise comparisons for several pairs of the efficient estimators, TMLE, C-TMLE, and DR-IPW, with different PS estimators. The purpose of these pairwise comparisons is to help in understanding the contribution of the collaborative estimation of the PS. We  used the shape and color of the points to represent the coverage information of the CIs for each estimates. 

\subsubsection{Impact of Collaborative  Propensity Score Model Selection}

We first compared the two pairs. Within the pair, both of the estimators were identical except each had a  different PS estimator. The first pair compared TMLE to TMLE*, and the second pair compared C-TMLE0 to CTMLE0*.

\begin{figure}[ht]
  \centering
  \begin{subfigure}[b]{0.35\textwidth}
    \includegraphics[width=\textwidth]{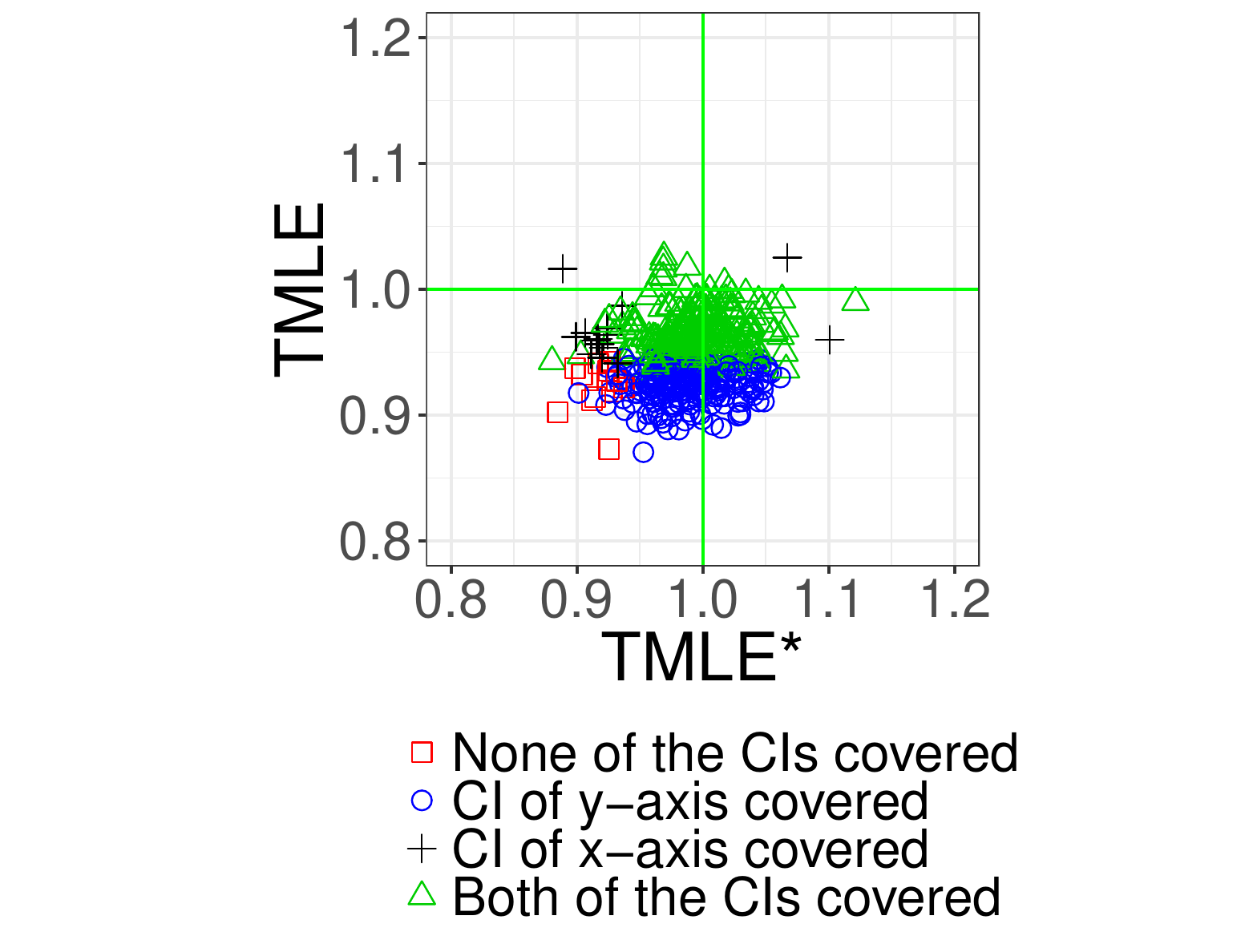}
    \caption{Comparison of TMLE and TMLE*, with the initial estimate $Q_n^0$ adjusting for 10 out of 40 confounders.}
    \label{fig:comp1_10}
  \end{subfigure}
  \hspace{2mm}
  \begin{subfigure}[b]{0.35\textwidth}
    \includegraphics[width=\textwidth]{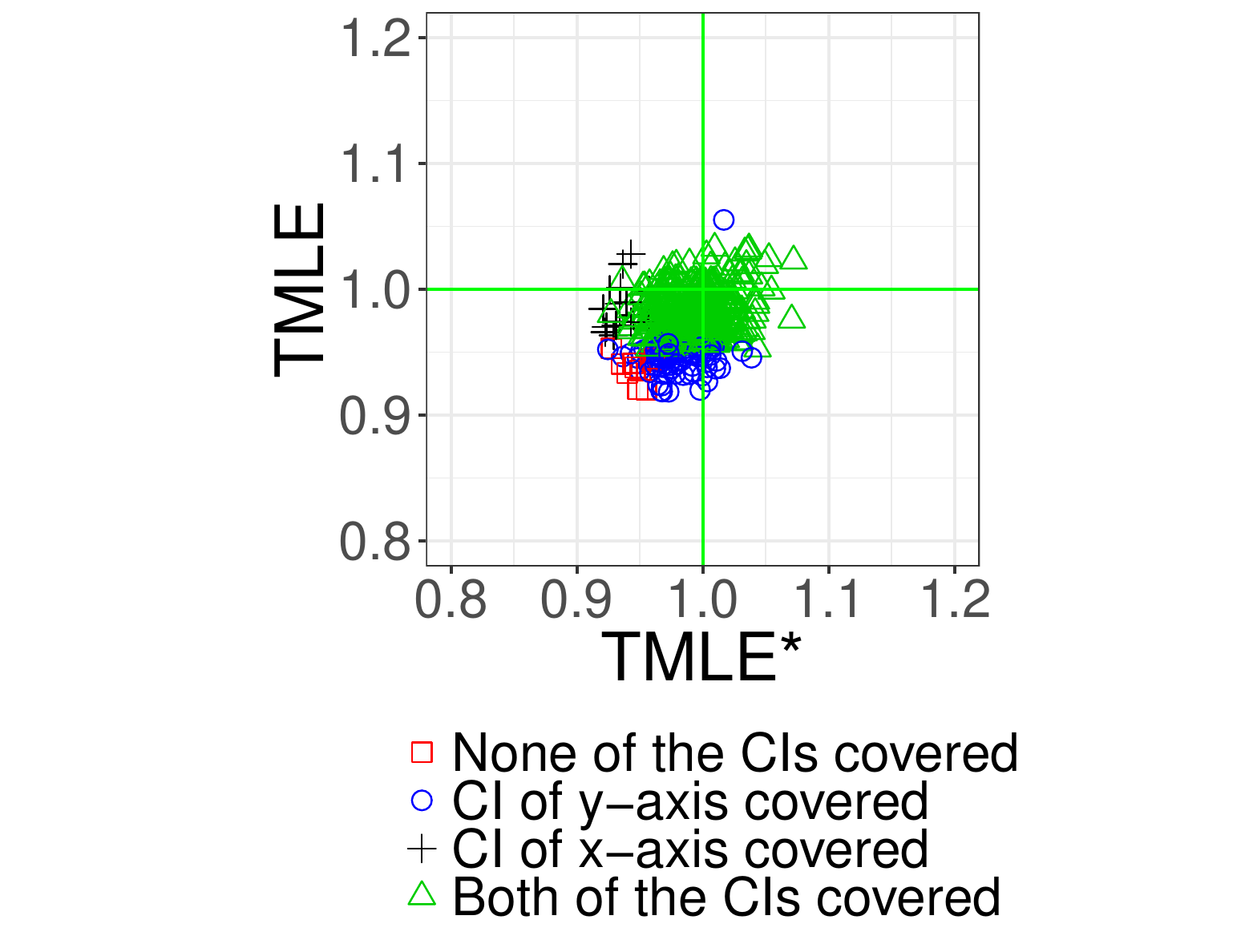}
    \caption{Comparison of TMLE and TMLE*, with the initial estimate $Q_n^0$ adjusting for 20 out of 40 confounders.}
    \label{fig:comp1_20}
  \end{subfigure}
  \caption{Comparison of TMLE wand TMLE*. The only difference within the pair the how the estimator $g_n$ is selected}
  \label{fig:comp_g1}
\end{figure}

From figure \ref{fig:comp1_10} and \ref{fig:comp1_20}, we can see that a more targeted PS model contributes substantially to the estimation. The vanilla TMLE underestimated the ATE, while TMLE* is close to unbiased. The variance of the two estimators are similar.

\begin{figure}[ht]
  \centering
  \begin{subfigure}[b]{0.35\textwidth}
    \includegraphics[width=\textwidth]{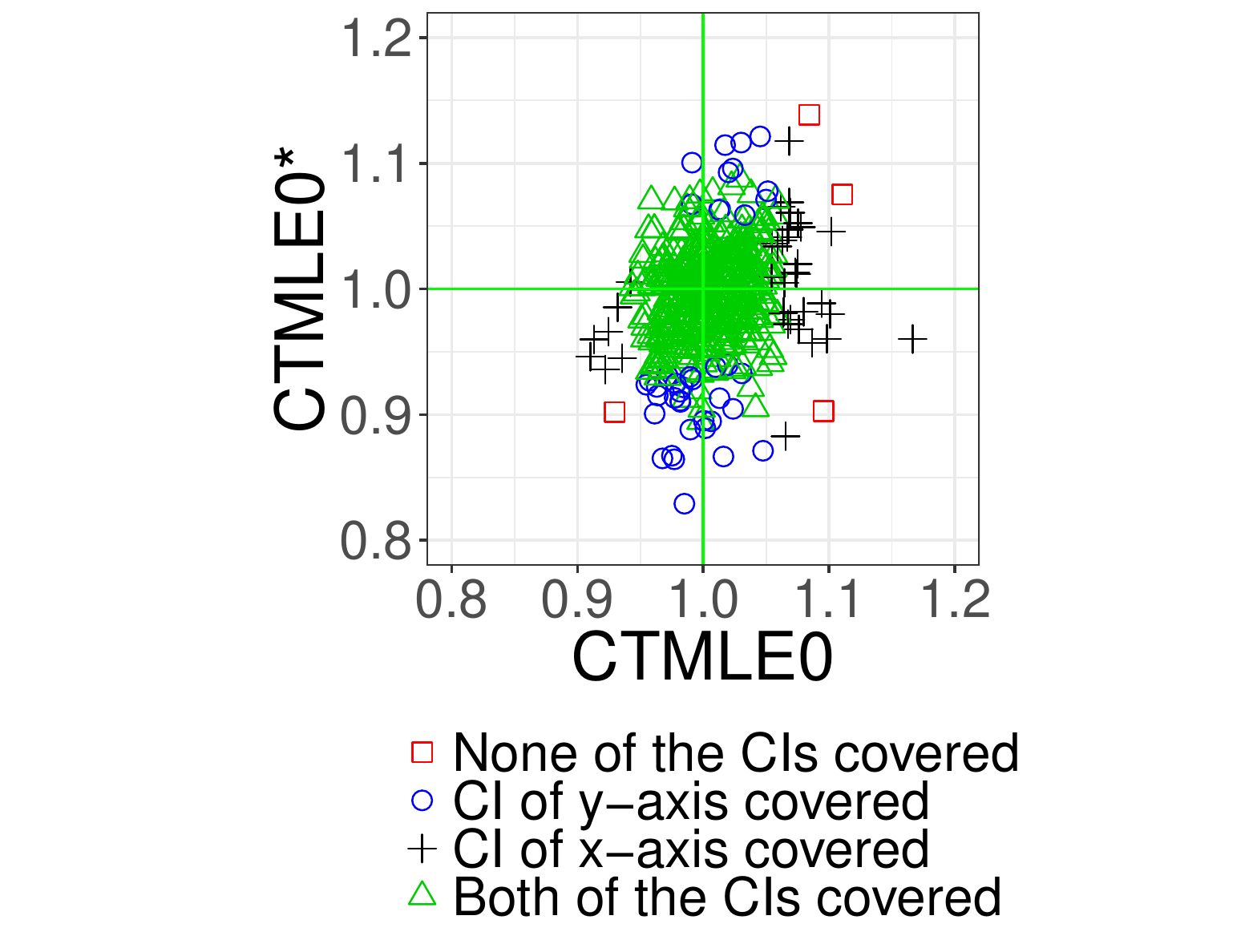}
    \caption{Comparison of C-TMLE0 and C-TMLE0*, with the initial estimate $Q_n^0$ adjusting for 10 out of 40 confounders.}
    \label{fig:comp3_10}
  \end{subfigure}\hspace{2mm}
  \begin{subfigure}[b]{0.35\textwidth}
    \includegraphics[width=\textwidth]{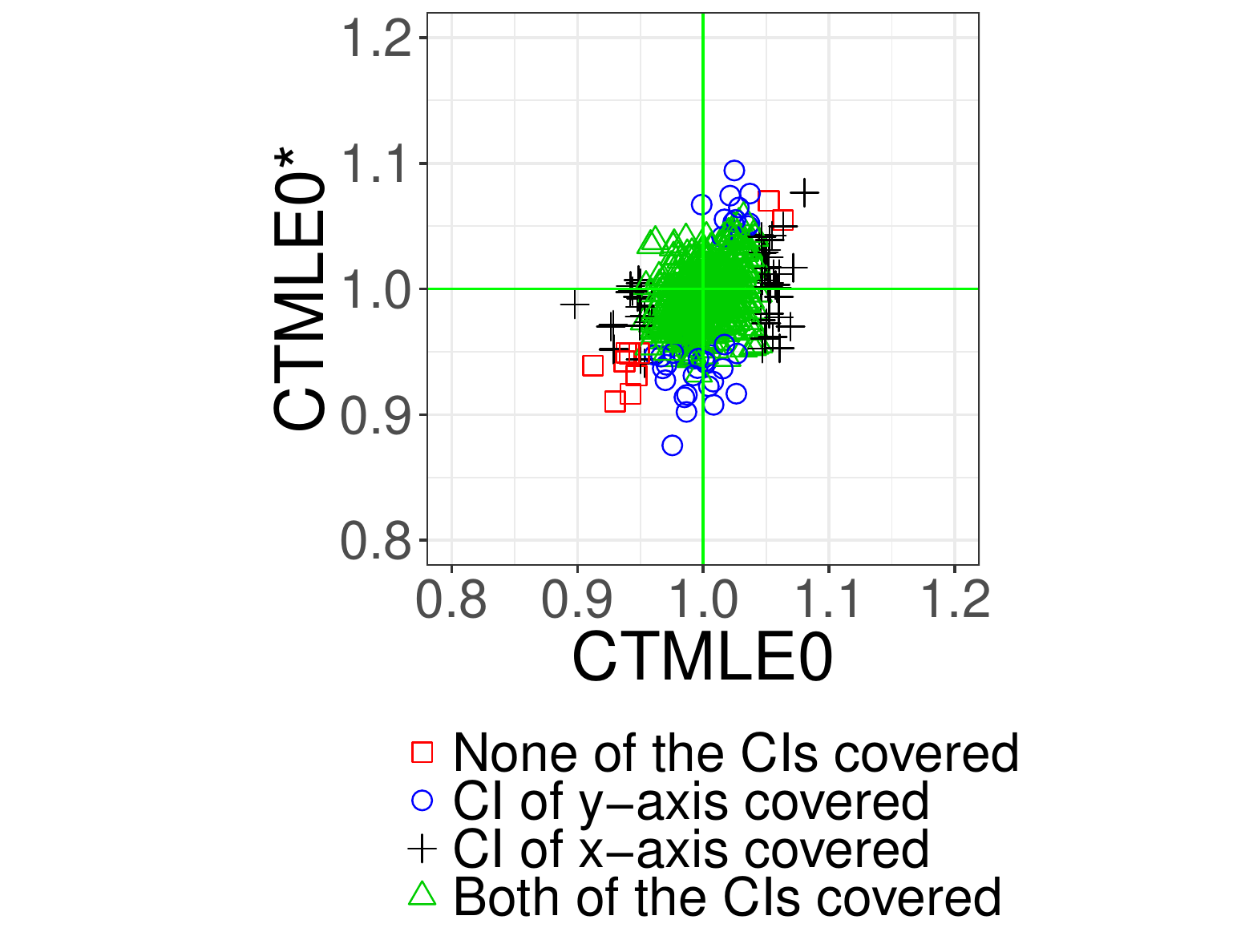}
    \caption{Comparison of C-TMLE0 and C-TMLE0*, with the initial estimate $Q_n^0$ adjusting for 20 out of 40 confounders.}
    \label{fig:comp3_20}
  \end{subfigure}
  \caption{Comparison of  CTMLE0 and CTMLE0*. The only difference within the pair the how the estimator $g_n$ is selected}
  \label{fig:comp_g1_2}
\end{figure}

From figure \ref{fig:comp3_10} and \ref{fig:comp3_20} we can see that the improvement for the CTMLE0 pair is not as significant as the improvement for the  TMLE pair. Interestingly, most of the poor performance in the CIs for CTMLE0 is from the over-estimated point estimate, while for CTMLE0* is mainly from under-estimation of the point estimate.

As discussed in \citep{ju2017ctmle,van2017ctmle}, such ignorable improvement with collaboratively selecting $g_n$ for the CTMLE0 pair  might be due to the redundant collaborative estimation step.  Thus, it is not necessary to both select the PS model using C-TMLE and solve for the extra clever covariate.


\subsubsection{Contribution of Solving Extra Critical Equation}

We compared TMLE with C-TMLE0. The only difference between these two estimators is that C-TMLE0 solves for the extra clever covariate, which guarantees that the critical equation is solved.

\begin{figure}[ht]
  \centering
  \begin{subfigure}[b]{0.35\textwidth}
    \includegraphics[width=\textwidth]{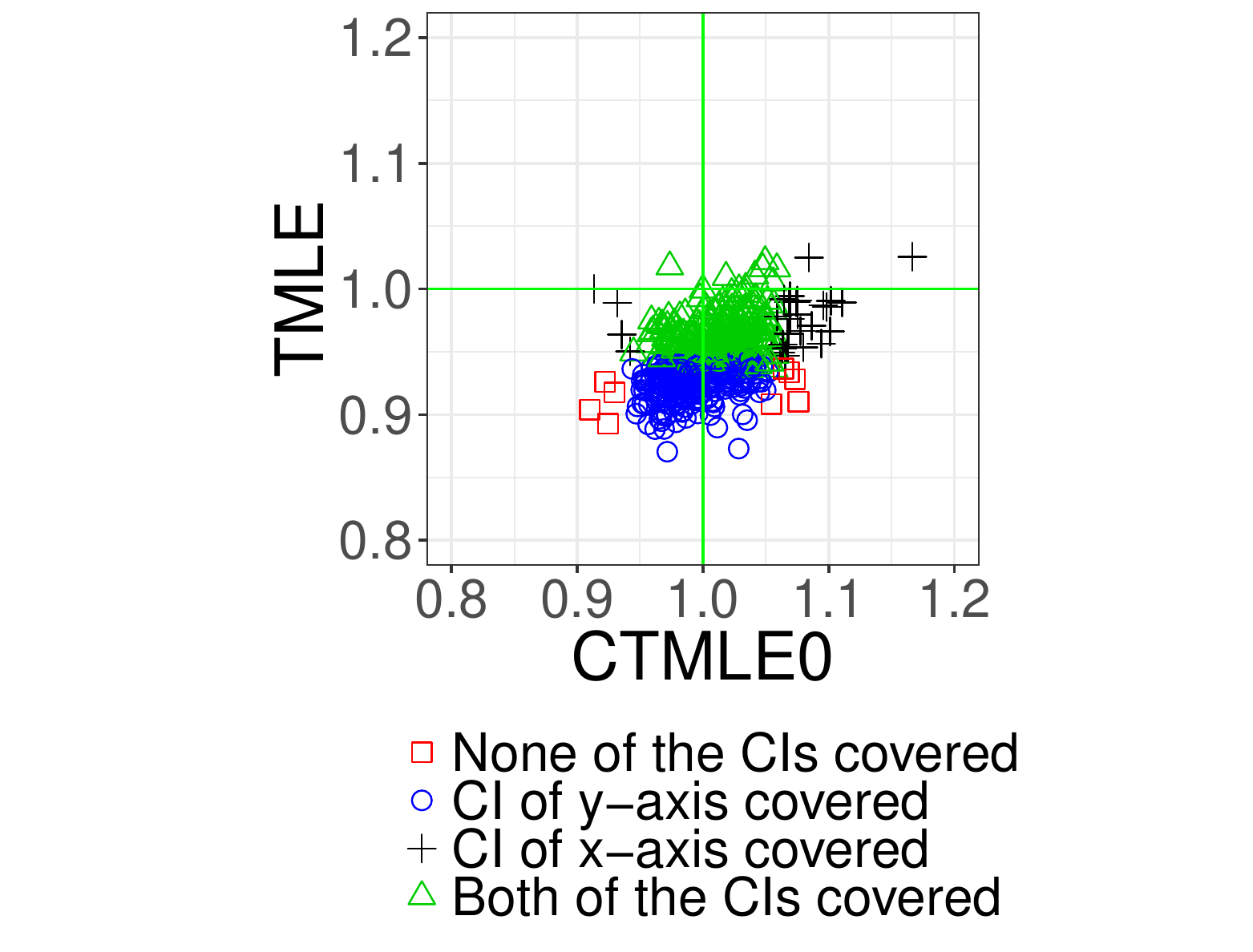}
    \caption{Comparison of TMLE and C-TMLE0, with the initial estimate $Q_n^0$ adjusting for 10 out of 40 confounders.}
    \label{fig:comp4_10}
  \end{subfigure}\hspace{2mm}
  \begin{subfigure}[b]{0.35\textwidth}
    \includegraphics[width=\textwidth]{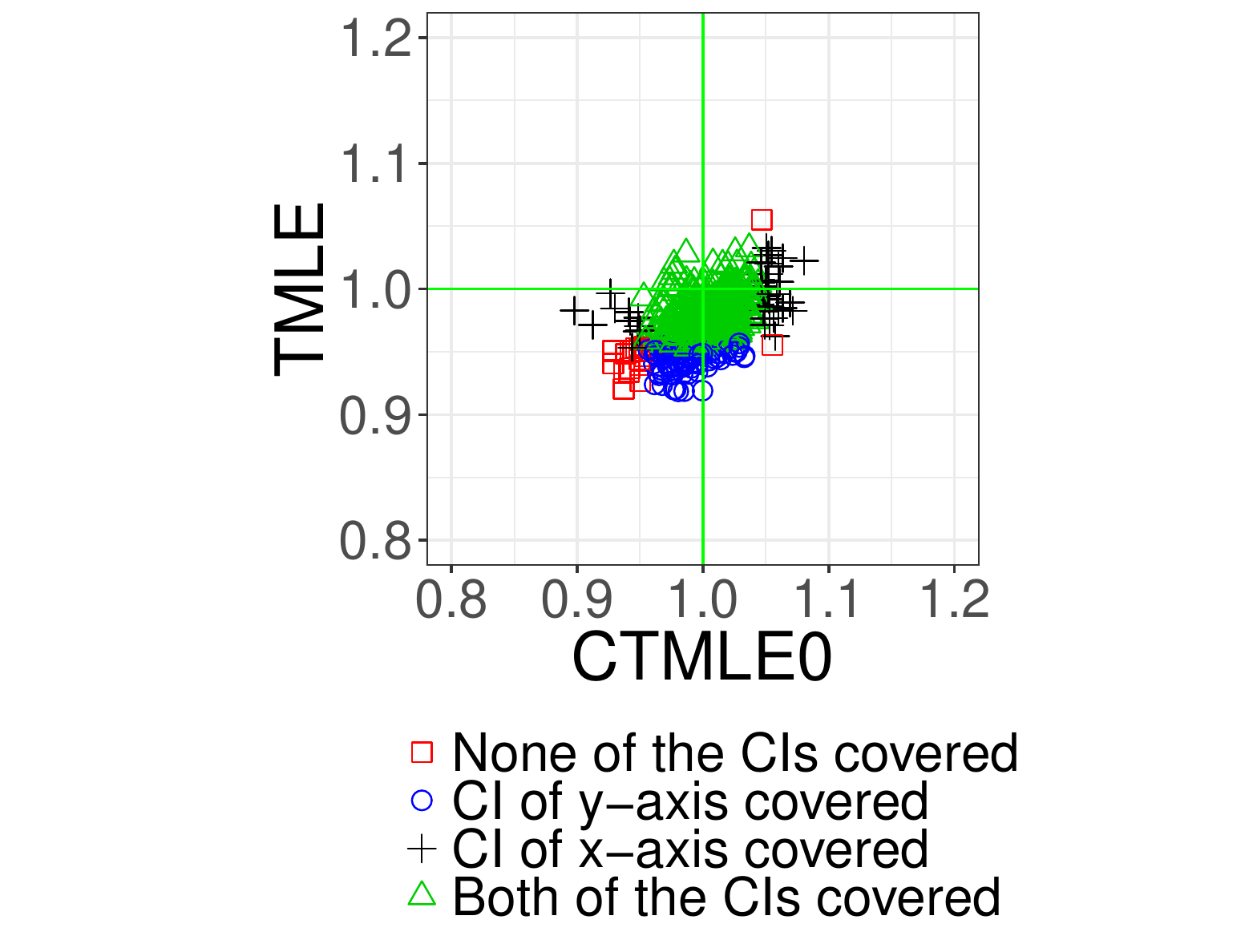}
    \caption{Comparison of TMLE and C-TMLE0, with the initial estimate $Q_n^0$ adjusting for 20 out of 40 confounders.}
    \label{fig:comp4_20}
  \end{subfigure}
  \caption{We compared TMLE with C-TMLE0, where the only difference between the two estimators is that C-TMLE0 solves the extra critical equation with additional clever covariates.}
  \label{fig:equation}
\end{figure}

Figure \ref{fig:equation} shows the improvement of solving an additional clever covariate. C-TMLE0 is less biased compared with TMLE. It is interesting to see that the performance of the estimator can improve substantially  with such small change. In addition, this additional change almost requires no additional computation, which makes it more favorable among proposed C-TMLEs when the computation resources are limited.

\subsubsection{Comparison of Variations of C-TMLE}

We compared the two pairs of variations of C-TMLEs. We used C-TMLE1 as the benchmark, as it gave the best performance for both point estimation and confidence interval coverage.

\begin{figure}[ht]
  \centering
  \begin{subfigure}[b]{0.35\textwidth}
    \includegraphics[width=\textwidth]{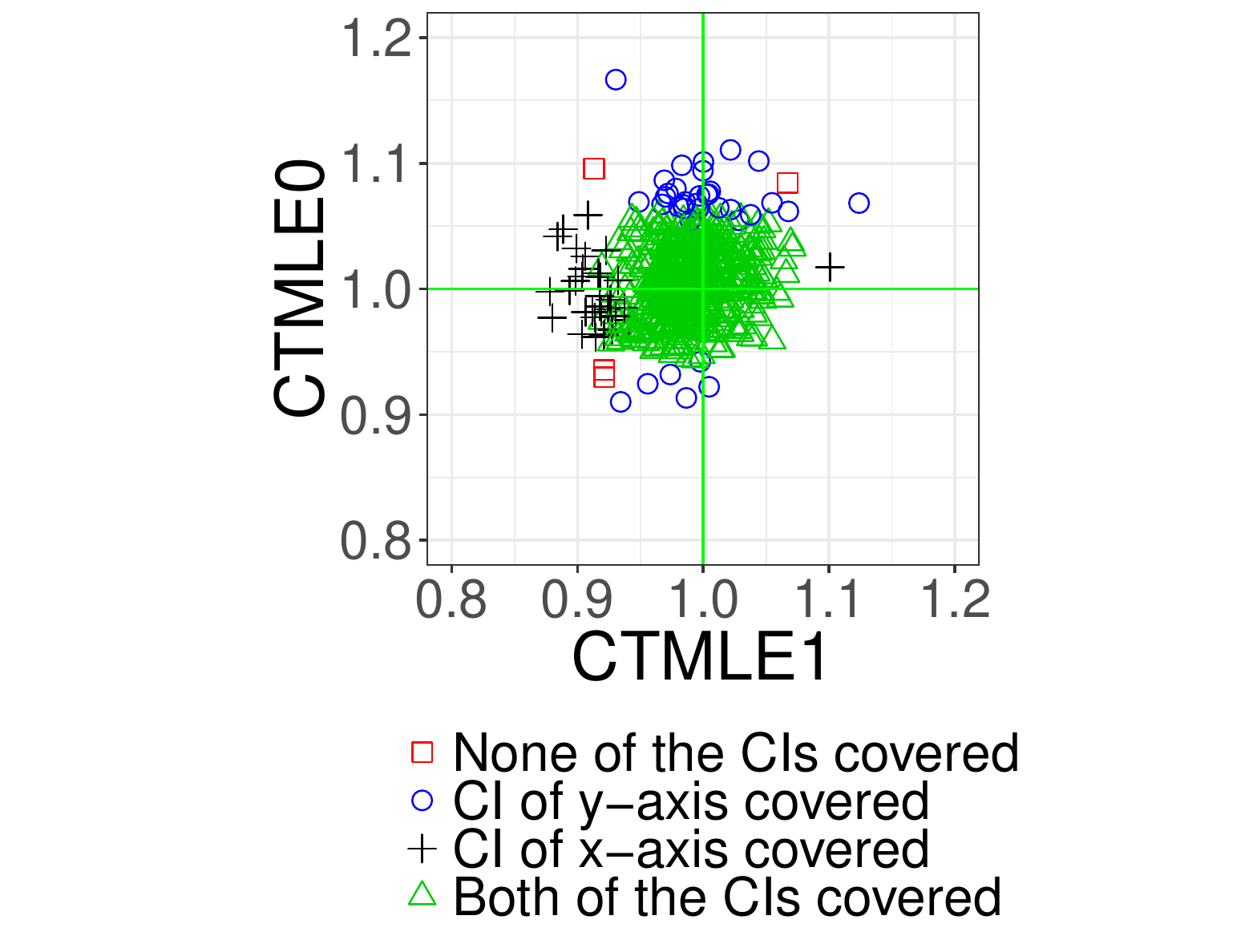}
    \caption{Comparison of C-TMLE1 and C-TMLE0, with the initial estimate $Q_n^0$ adjusting for 10 out of 40 confounders.}
    \label{fig:comp6_10}
  \end{subfigure}\hspace{2mm}
  \begin{subfigure}[b]{0.35\textwidth}
    \includegraphics[width=\textwidth]{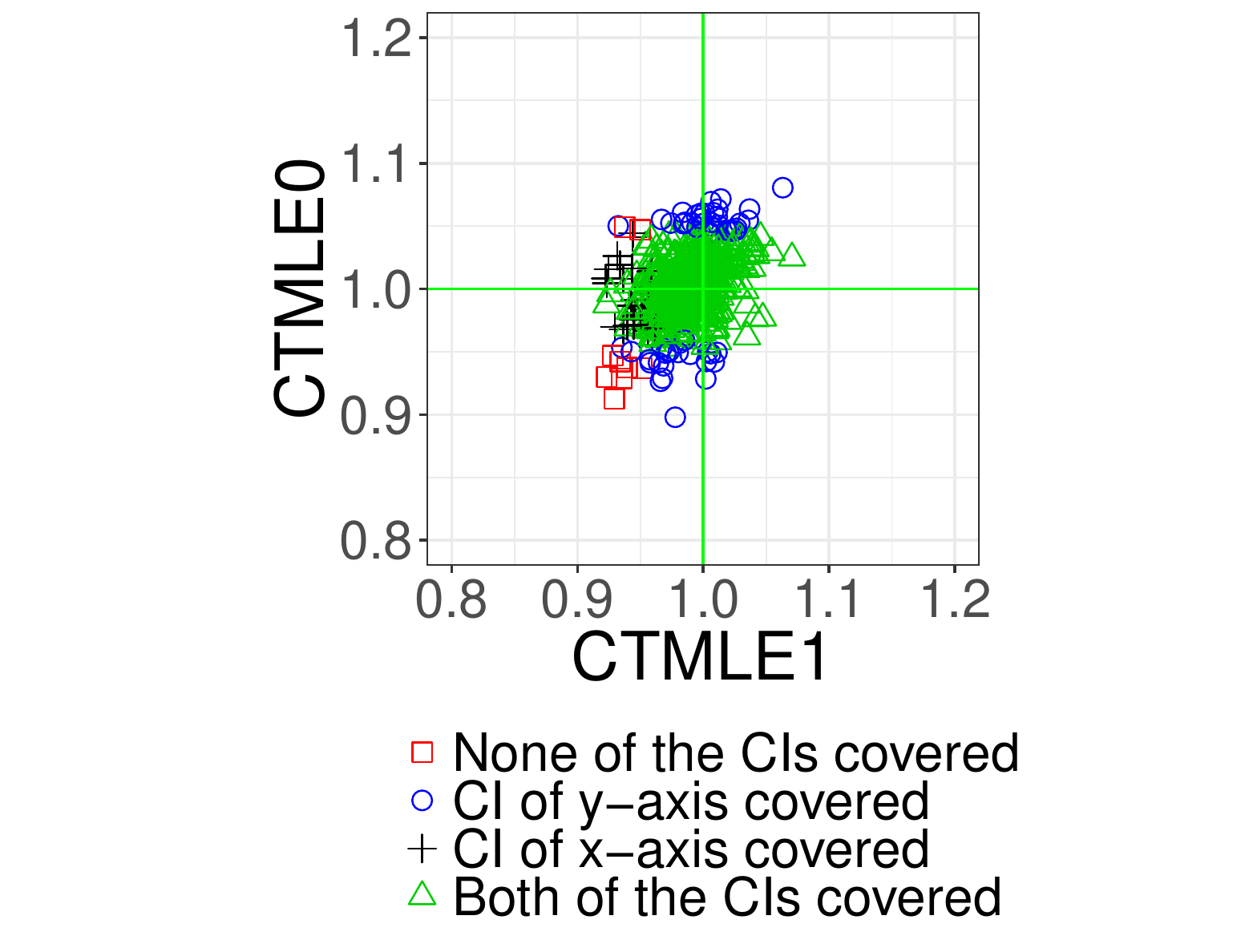}
    \caption{Comparison of C-TMLE1 and C-TMLE0, with the initial estimate $Q_n^0$ adjusting for
     20 out of 40 confounders.}
    \label{fig:comp6_20}
  \end{subfigure}
  \caption{We compared C-TMLE1 with C-TMLE0.}
  \label{fig:comp_ctmles}
\end{figure}

Figure \ref{fig:comp6_10} and \ref{fig:comp6_20} show the pairwise performance of C-TMLE1 and C-TMLE0. Both estimators performed well with respect to the MSE. Although the distribution of points looks similar and have variances that appear similar, there were more CIs from C-TMLE0 that failed to cover the truth. In addition, the failures from C-TMLE1 mainly resulted from the under-estimation of the estimates. In comparison, the failures from C-TMLE0 primarily came from both under/over-estimated estimates. This suggests that the relatively poor CI coverage of C-TMLE0 might be due to its under-estimated standard error.

\section{Real Data Analysis}
\label{sec:da}

In this section, we applied the methods described previously to  the NSAID study. As discussed previously, the goal of this study is to compare the effectiveness of two treatments  on improving the risk (probability) of being diagnosed with severe gastrointestinal complications during the follow-up period. The treatment group was prescribed  a selective COX-2 inhibitor, while the control group was prescribed a non-selective nonsteroidal anti-inflammatory drug. To compare the safety of the two treatments, we used the average treatment effect (ATE) as our target parameter.

\subsection{Method}

We followed the hdPS procedure in subsection \ref{subsec:hdps}, where we generated the hdPS covariates with $k_1 = 100$ and $k_2 = 200$.

  We investigated three kinds of initial estimate $\bar{Q}_n^0$ for TMLE and C-TMLE:

  \begin{itemize}

  \item The initial estimate was given by the group means of the treatment and control group.

  \item The initial estimate was estimated by Super Learner with only baseline covariates.

  \item The initial estimate was estimated by Super Learner with both baseline covariates and hdPS covariates.

  \end{itemize}
   For Super Learners \citep{van2007super,polley2010super}, we used library with LASSO \citep{friedman2009glmnet}, Gradient Boosting Machine \citep{ridgeway2006gbm}, and Extreme Gradient Boosting \citep{chen2015xgboost}.

\subsection{Results}

\begin{figure}[ht]
  \centering
    \begin{subfigure}[t]{0.3\textwidth}
    \includegraphics[width=\textwidth]{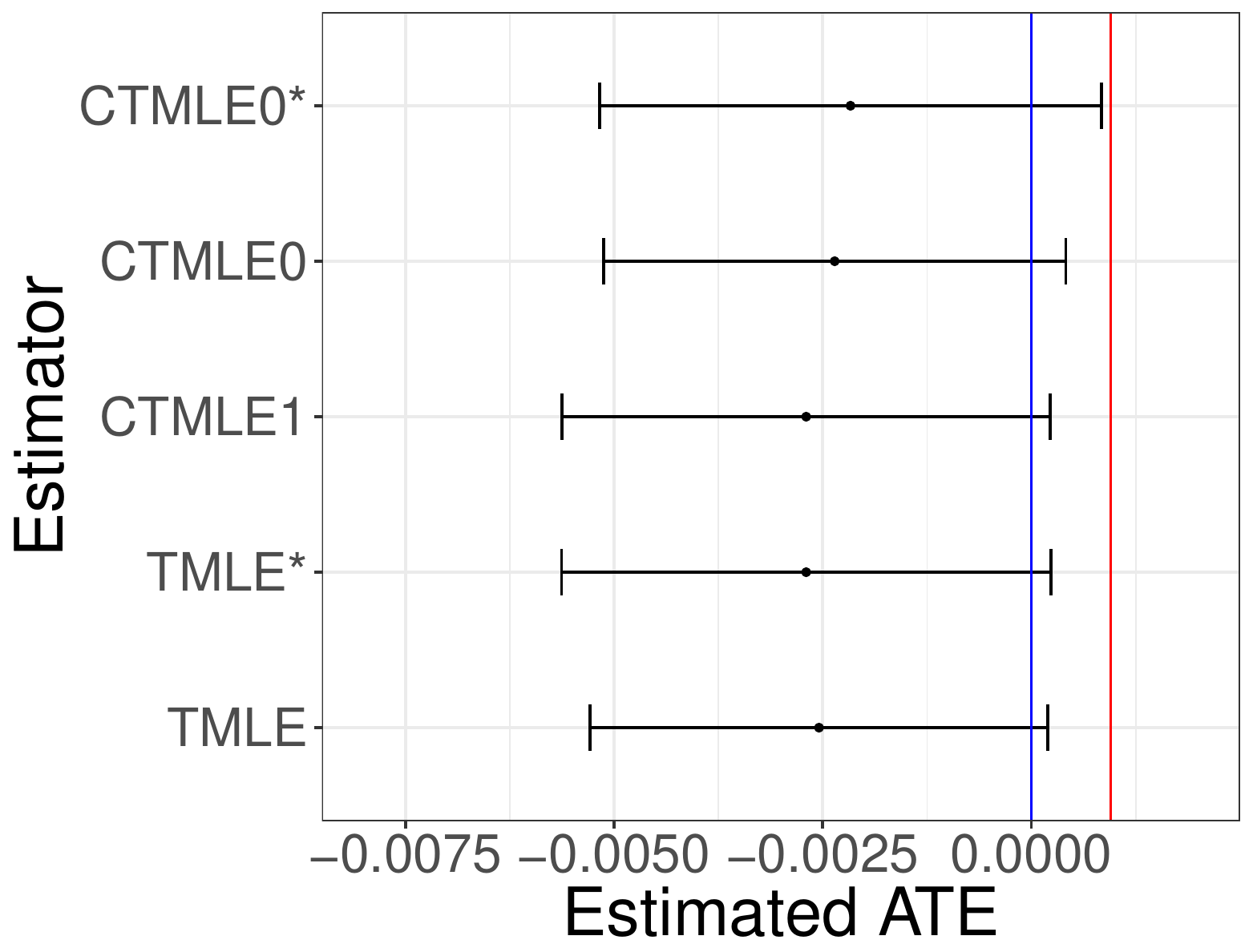}
    \caption{Influence curve based confidence interval for all TMLE based estimators for NSAID study, with the group means as initial estimate.}
    \label{fig:nsaid_ci_ic_intercept}
  \end{subfigure}
  \hspace{5mm}
  \begin{subfigure}[t]{0.3\textwidth}
    \includegraphics[width=\textwidth]{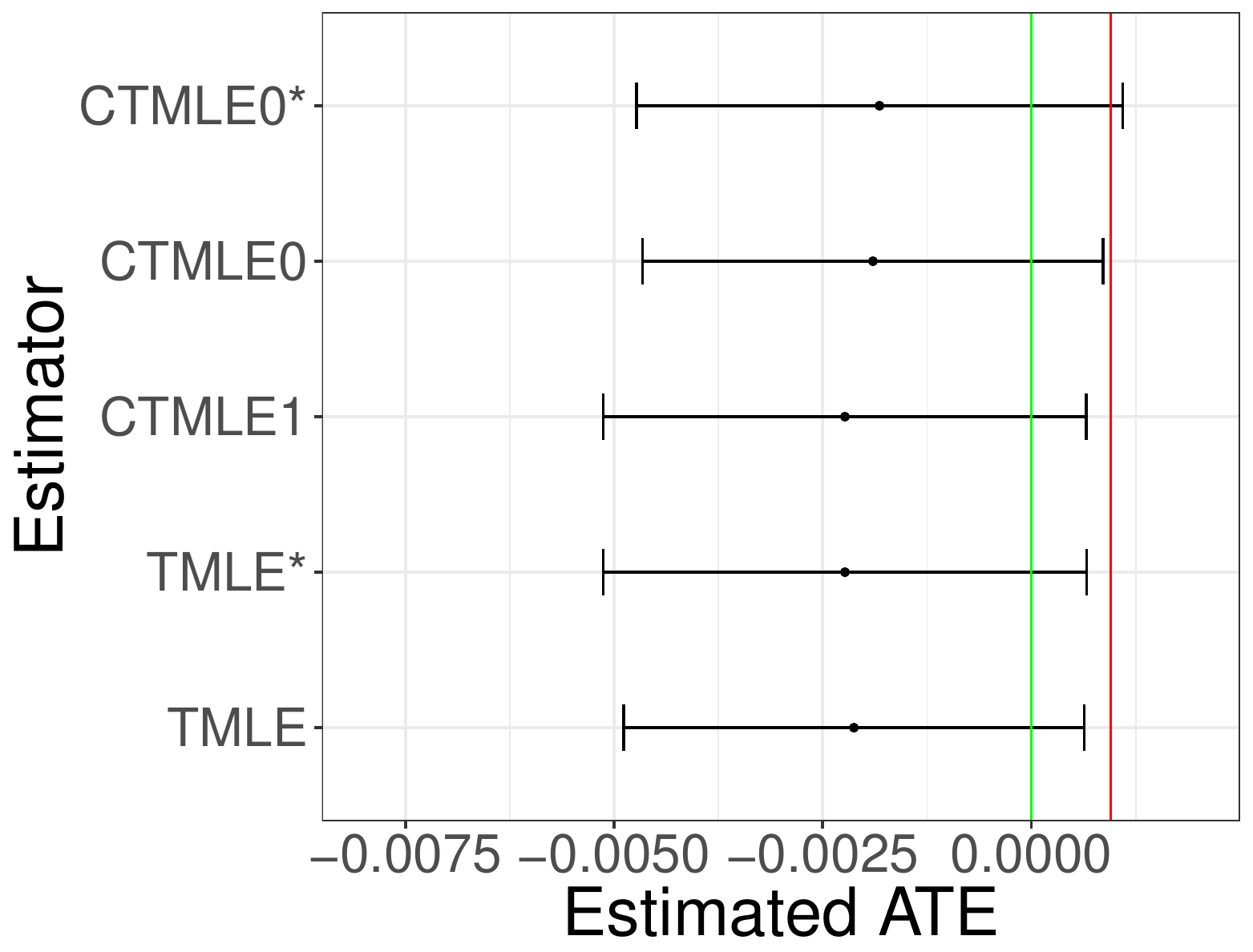}
    \caption{Influence curve based confidence interval for all TMLE based estimators for NSAID study, with initial estimate provided by Super Learner with baseline covariates.}
    \label{fig:nsaid_ci_ic_main}
  \end{subfigure}
  \hspace{5mm}
  \begin{subfigure}[t]{0.3\textwidth}
    \includegraphics[width=\textwidth]{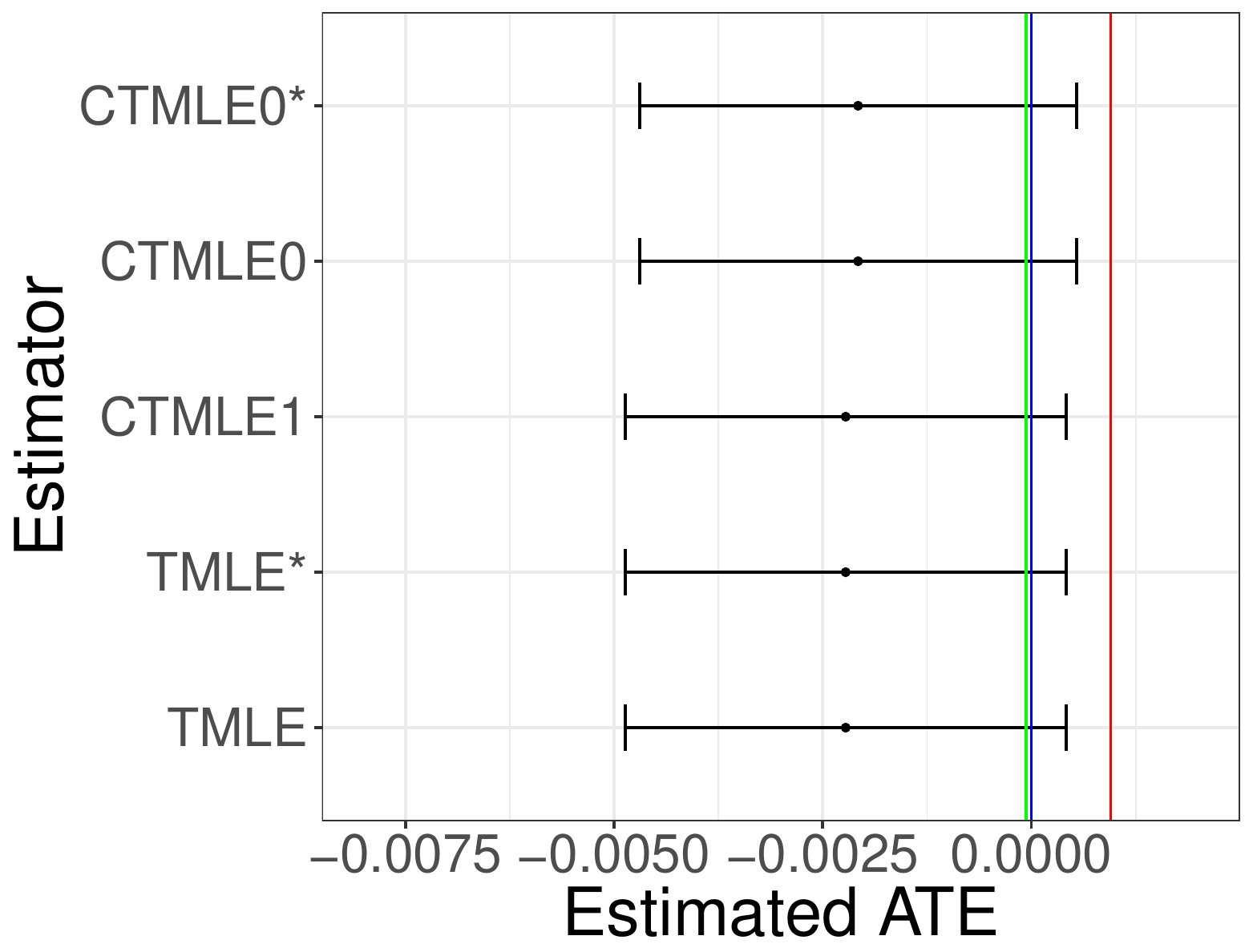}
    \caption{Influence curve based confidence interval for all TMLE based estimators for NSAID study, with the initial estimate provided by Super Learner with baseline covariates and hdPS covariates.}
    \label{fig:nsaid_ci_ic_full}
  \end{subfigure}
  \caption{Confidence intervals for TMLE based estimators for the NSAID study. }
  \label{fig:nsaid_cis}
\end{figure}

Figure \ref{fig:nsaid_cis} shows the point estimates and $95\%$ CIs for all TMLE and C-TMLE estimators. We use the blue line to denote the null hypothesis ($\text{H}_0:\Psi_0 = 0$),  the green line denotes the initial estimate, 
and use red line to denote the results from the naive difference in means estimator ($\Psi_n^{\text{naive}} = 0.0949\%$).

Figure \ref{fig:nsaid_ci_ic_full} shows that, after adjusting for selection bias using the TMLE/C-TMLE algorithms, all the estimators have similar results, with the  estimated ATE being in the negative direction.  Similar to the results in simulation, the CIs for TMLE* and C-TMLE0* were wider with PS estimator selected by C-TMLE1, than with PS estimator selected by CV. The details of the  point estimates and confidence intervals are reported in table~\ref{tab:nsaid}. We  computed the analytic influence curve based confidence interval. None of these intervals, except C-TMLE0*, covered the naive estimate. However, all of them covered the null hypothesis.

\begin{table*}
  \centering
  \caption{The point estimates for all TMLE/C-TMLE estimators. All the values are on a scale of $10^{-2}$. \label{tab:nsaid}}
  \begin{tabular}{|r|rrrrr|}
    \hline
names & TMLE & TMLE* & CTMLE1 & CTMLE0 & CTMLE0* \\  \hline
Point Estimate & -0.2381 & -0.2491 & -0.2491 & -0.2208 & -0.2093 \\ 
  Estimated SE & 0.1414 & 0.1487 & 0.1486  & 0.1417 & 0.1502 \\ 
   \hline
\end{tabular}
\end{table*}

In addition, we also compared the results from different initial estimator. Figure \ref{fig:nsaid_cis} shows the results for all estimators, with group means (\ref{fig:nsaid_ci_ic_intercept}), Super Learner with baseline covariates (\ref{fig:nsaid_ci_ic_main}), and Super Learner with both baseline and hdPS covariates (\ref{fig:nsaid_ci_ic_full}). The CV.LASSO PS estimator selected 137 covariates, with regularization parameter $\lambda = 0.001159$. The C-TMLE estimator with naive initial estimate selected $164$ covariates, with $\lambda = 0.000266$. The C-TMLE estimator uses the initial estimate provided by SL with only baseline covariate have similar results: it selected $166$ covariates with $\lambda=0.000238$. For the C-TMLE with initial estimate provided by SL with all covariates, it selected the same model as CV.LASSO. It shows when the initial estimate is biased, C-TMLE selected model with less regularization, thus adjusted more potential confounders. In addition, all the covariates that included by LASSO selected by C-TMLE but not by CV.LASSO are hdPS covariates. This suggests such additional hdPS covariates can be confounder. However, as they have relatively weaker predictive performance for treatment mechanism, they would be mistakenly removed by CV.LASSO.

\begin{figure}[ht]
  \centering
    \includegraphics[width=0.6\textwidth]{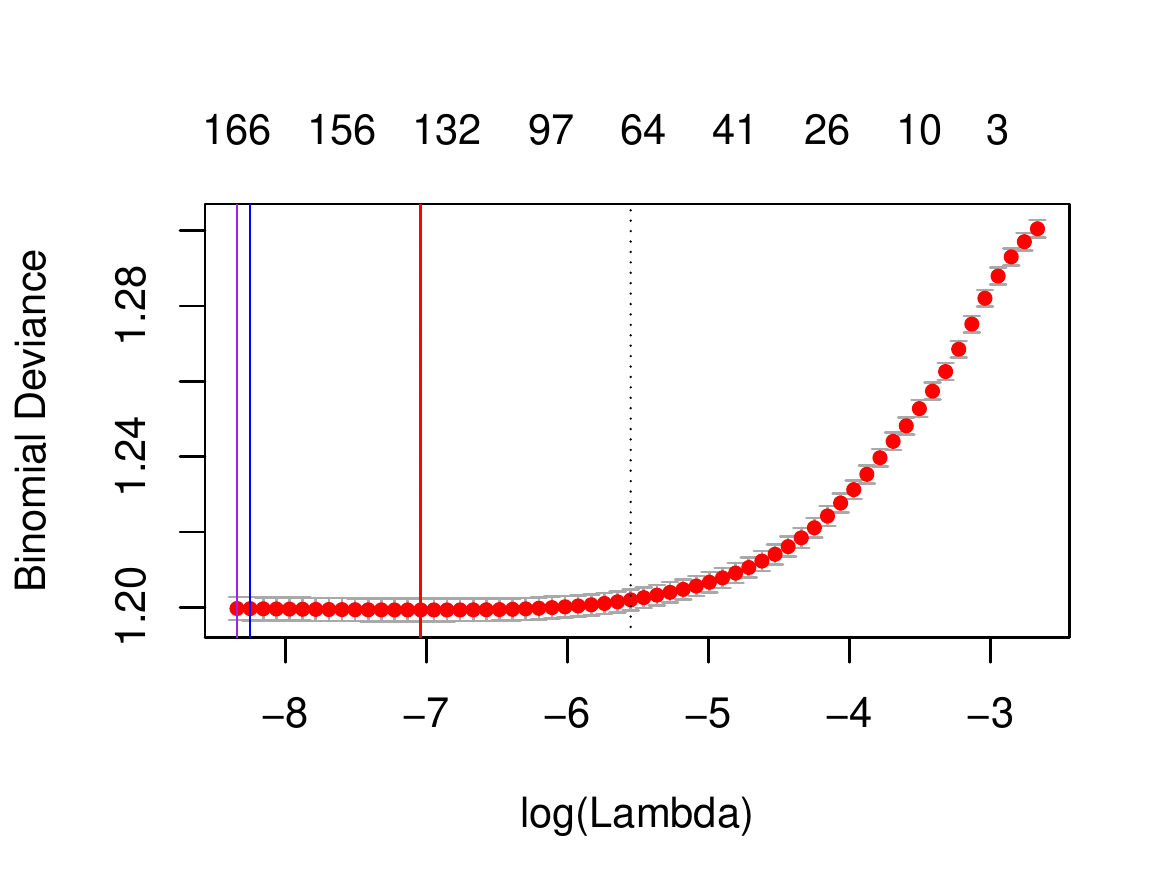}
  \caption{Binomial deviance for $\lambda$ selected by CV.LASSO and C-TMLE with different initial estimators.\label{fig:deviance}}
  
\end{figure}

 Figure \ref{fig:deviance} shows the details of the CV loss for each selected PS estimator. The blue line is the $\lambda$ selected by C-TMLE1 with naive  estimator. Its CV binomial deviance (twice the binomial negative log-likelihood) is  $1.199632$. The purple line is the  $\lambda$ selected by C-TMLE1 with initial estimator provided by SL with only baseline covariates. Its CV binomial deviance is $1.199668$. The red line is the $\lambda$ selected by CV.LASSO, and C-TMLE1 with initial estimator provided by SL with both baseline and hdPS covariates. Its CV binomial deviance is $1.199288$.
 This may be due to the signals in all the initial estimates are too weak: all the initial estimates of ATE are very close to 0. In addition, all the confidence intervals covered null hypothesis. The additive treatment effect in this study is not statistically significant.

\subsection{Conclusions from the Empirical Study}

Patients who received selective COX-2 inhibitors were less likely to get  severe gastrointestinal complications  during the follow-up period, compared to the patients who received a non- selective nonsteroidal anti-inflammatory drug. The average additive treatment effect was approximately $-0.249\%$, which was estimated using  TMLE* and C-TMLE1 (the two estimators achieved the best performance in simulations).  The point estimates for other estimators were similar.

Based on the results, the additive treatment effect was not statistically significant. However, this does not necessary imply that there is no difference between the two treatments. More observations or better designed studies are necessary for further comparison of these treatments.

\section{Conclusion}
\label{sec:diss}

In this study, we described two variations of C-TMLE, and assessed their performance on quasi-experiments based on real empirical data. We assessed the performance of several well studied PS-based estimators  in settings where estimated models for  both the conditional response $\E(Y|A, W)$ and the propensity score $\E(A|W)$  were misspecified. In particular, we focused on using the LASSO estimator  for the PS model. In comparison to our previous work, this study provides a more detailed evaluation of all the estimators by not only assessing their point estimation, but also the confidence intervals  for each of the estimators. Results showed that the C-TMLE1 and C-TMLE0 estimators had the best performance in terms of both point estimation and CI. We also evaluated the impact of directly applying the model that was  collaboratively selected by C-TMLE1 to other PS non-collaborative estimators. Results showed that all of the PS-based estimators, except the vanilla IPW estimator, improved substantially, in terms of the point estimation, when the collaboratively selected model was applied to these estimators. However, C-TMLE0* did not improve when compared to C-TMLE0 for point estimation. Finally, pairwise comparisons of estimators were also evaluated to help in understanding the contribution of the collaborative model selection.

In comparison to previous work, this study is the first to thoroughly investigate and compare the confidence intervals coverage and length for the novel C-TMLE algorithms, as well as some commonly used competitors. Further, it offers detailed pair-wise comparisons with other competing estimators using different PS model selection procedures. Finally, this study utilizes the quasi-experiments based on a real electronic healthcare dataset and then makes inference on the same database. This makes the conclusions from the real data analysis more convincing.

In conclusion, this study introduces a new direction for PS model selection. It shows the insufficiency of using ``external'' cross-validation for the LASSO estimator. Thus, we conclude that the ensemble PS estimators, which rely on ``external'' cross-validation, are not  optimal (w.r.t. the causal parameter) for maximizing confounding control. Ensemble learning that is based on C-TMLE is a potential solution to address this issue. We leave this for the future work.


\newpage
\bibliographystyle{abbrvnat}
\bibliography{references}

\newpage
\appendix

\begin{center}
  {\huge\bf Appendix}
\end{center}

\section{C-TMLE 1}
\label{sec:ctmle1}

C-TMLE1 is a straightforward instantiation of the general C-TMLE template, which generates a sequence of PS estimators, with corresponding TMLE estimators. Then it selects the TMLE estimator with the smallest cross-validated loss w.r.t. the causal parameter. Finally it takes one more targeting step to make sure the critical equation \ref{eq:critical} is solved \citep{ju2017ctmle,van2017ctmle}. Algorithm \ref{algo:ctmle1} shows the details of the C-TMLE1 algorithm.

\begin{algorithm}[H]
  \caption{Collaborative Targeted Maximum Likelihood Estimation Algorithm I \label{algo:ctmle1}}
  \begin{algorithmic}[1]
    \State{Construct an initial estimate $\bar{Q}_n^0$ for $\bar{Q}_0 =
      \E_0(Y \mid A, W).$ } \State{Construct a sequence of propensity score model
      $g_{n,\lambda}$ indexed by $\lambda$, where a larger $\lambda$ implies a
      smoother estimator (e.g. larger regularization for LASSO, or larger
      bandwidth for kernel estimator). We further set $\lambda$ within the set
      $\Lambda = [\lambda_{\min}, \lambda_{cv}]$.} \State{Bound the estimated
      propensity score $g_{n,\lambda} = \max\{0.025, \min\{g_{n,\lambda},
      0.975\}\}$}

    \State{Set $k = 0$}
    \While{$\Lambda$ is not empty}
    \State{Apply targeting
      step for each $g_{n,\lambda}$, with $\lambda \in \Lambda$, with initial estimate
      $\bar{Q}_n^k$ and clever covariate $H_{g_{n,\lambda}}(A,W)= \frac{1-A}{1
        -g_{n,\lambda}(W) } + \frac{A}{g_{n,\lambda}(W)}$.}
    \State{Select
      $\bar{Q}_{n, \lambda_k}^*$ with the smallest empirical risk $L(\bar{Q}_{n,
        \lambda_k}^*(A,W))$.}
    \State{For $\lambda \in [\lambda_{k}, \lambda_{k-1}]$,
      compute the corresponding TMLE using initial estimate $\bar{Q}_n^{k-1}$ and
      propensity score estimate $g_{n,\lambda}$. We denote such estimate with
      $\bar{Q}_{n,\lambda}^*$ and record them.}
    \State{Set a new initial estimate $\bar{Q}_n^{k} = \bar{Q}^*_{n, \lambda_k}$.}
    \State{Set $\Lambda =
      [\lambda_{\min}, \lambda_k)$.}
      \State{Set $k = k + 1$.}
      \EndWhile
      \State{Select the best candidate $
        \bar{Q}_{n,\lambda_{ctmle}}^*$ among $\bar{Q}_{n,\lambda}^*$, with the smallest
        cross-validated loss, using the same loss function as in the TMLE targeting
        step.}
      \State{Pick up the corresponding initial estimate $\bar{Q}_{n,\lambda_{ctmle}}$ for $\bar{Q}^*_{n,\lambda_{ctmle}}$}
      \State{Apply targeting step to $\bar{Q}_{n,\lambda_{ctmle}}$ from the last step, with each $g_{n,\lambda}$, $\lambda \in [\lambda_{min},\lambda_{ctmle})$, yielding a new sequence of estimate $\bar{Q}_{n,\lambda}^*$.}
        \State{Select $\bar{Q}_n^{*} = \argmin_{\bar{Q}_{n,\lambda}^{*}}L(\bar{Q}_{n,\lambda}^{*}), \lambda \in [\lambda_{min},\lambda_{ctmle})$  with the smallest empirical loss from the sequence in last step as the final estimate.}
  \end{algorithmic}
\end{algorithm}

\section{C-TMLE 0}
\label{sec:ctmle3}


In the C-TMLE0 algorithm, we only fluctuate the initial estimate using two clever covariates,
$H_{g_{n,\lambda}}(A,W)$ and $\tilde{H}_{g_{n,\lambda}}(A,W)$, with propensity score estimate $g_{n,\lambda}=g_{n, \lambda_{cv}}$ pre-selected by external cross-validation. 

One of the main strength of this method is its computational efficiency: without generating sequence of TMLE estimators and applying cross-validation for model selection, it is much faster compared to C-TMLE1. Algorithm \ref{ctmle3} shows the detail of the C-TMLE0 algorithm.

\begin{algorithm}[H]
  \caption{Collaborative Targeted Maximum Likelihood Estimation 0}
  \label{ctmle3}
  \begin{algorithmic}[1]
    \State{Construct an initial estimate $\bar{Q}_n^0$ for $\bar{Q}_0 =
      \E_0(Y \mid A, W).$ } \State{Estimate the propensity score and select the
    hyper-parameter using external cross-validation: $g_{n,\lambda}=g_{n, \lambda_{cv}}$.} 
    \State{Apply targeting step in \eqref{eq:target} with initial estimate $\bar{Q}_n^0$ and two
      clever covariates
      \begin{equation*}
    \begin{aligned}
      H_{g_{n,\lambda}}(A,W)= \frac{1-A}{1 -g_{n,\lambda}(W) } +
      \frac{A}{g_{n,\lambda}(W)}
    \end{aligned}
    \end{equation*}
    and
    \begin{equation*}
    \begin{aligned}
      \tilde{H}_{g_{n,\lambda}}= &-\frac{1-A}{(1 -g_{n,\lambda}(W))^2}(g_{n,\lambda + \delta} -
    g_{n,\lambda} ) \\&+ \frac{A}{g_{n,\lambda}(W)} (g_{n,\lambda + \delta} -
    g_{n,\lambda} ) ,
    \end{aligned}
    \end{equation*}
    which gives a new estimate $\bar{Q}_{n,\lambda_{cv}}^{*}$.}
 
    \State{Return the TMLE:
    $\bar{Q}_{n}^* = \bar{Q}_{n,\lambda_{cv}}^{*}$}
  \end{algorithmic}
\end{algorithm}

\end{document}